\documentclass[12pt]{article}
\RequirePackage{amsthm,amsmath,amsfonts,amssymb}
\RequirePackage[authoryear]{natbib}
\RequirePackage[colorlinks,citecolor=blue,urlcolor=blue]{hyperref}
\RequirePackage{graphicx}

\usepackage{graphicx}
\usepackage{multirow}
\usepackage{mathrsfs}
\usepackage{bigints}
\usepackage{bm}
\usepackage{bbm}
\usepackage{setspace}
\usepackage{xcolor}
\usepackage{verbatim}
\usepackage{caption}
\usepackage{subcaption}
\usepackage{slashbox}
\usepackage{natbib}
\usepackage{booktabs}
\usepackage{algorithm}
\usepackage{algpseudocode}
\usepackage{titling}
\usepackage[percent]{overpic}
\usepackage{float}
\usepackage{placeins} 

\usepackage{mac-09-ams-03-2021}

\newtheorem{theorem}{Theorem}[section]
\newtheorem{corollary}{Corollary}[theorem]
\newtheorem{lemma}[theorem]{Lemma}

\newtheorem{proposition}[theorem]{Proposition}

\newcommand{\blind}{1}

\begin{document}

	\def\spacingset#1{\renewcommand{\baselinestretch}%
		{#1}\small\normalsize} \spacingset{1}

	
	\if1\blind
	{
		\title{\bf Fitting Sparse Markov Models to Categorical Time Series Using Convex Clustering}
		\author{Tuhin Majumder\\
			Department of Biostatistics and Bioinformatics, Duke University\\
			and \\
			Soumendra Lahiri\thanks{
				This material is based upon work supported by the National Science Foundation under Grant DM 
				1811933.}\hspace{.2cm} \\
			Department of Mathematics and Statistics, Washington University in St. Louis\\
			and \\
			Donald Martin \\
			Department of Statistics, North Carolina State University}
		\maketitle
	} \fi
	
	\if0\blind
	{
		\bigskip
		\bigskip
		\bigskip
		\begin{center}
			{\LARGE\bf Fitting Sparse Markov Models to Categorical Time Series Using Convex Clustering}
		\end{center}
		\medskip
	} \fi
	
	\bigskip
	\begin{abstract}
		Higher-order Markov chains are frequently used to model categorical time series. However,  
		a major problem with fitting such models is the exponentially growing number of parameters in the model order. A popular approach to parsimonious modeling is to use a Variable Length Markov Chain (VLMC), which determines relevant contexts (recent pasts) of variable orders and forms a context tree. A more general parsimonious modeling approach is given by Sparse Markov Models (SMMs), where all possible histories of order $m$ are partitioned such that the transition probability vectors are identical for the histories belonging to any particular group. In this paper, we develop an elegant method of fitting SMMs  
		based on convex clustering and regularization. The regularization parameter is selected using the BIC criterion. Theoretical results establish model selection consistency of our method for large sample size.
		Extensive simulation results under different set-ups are presented to study 
		finite sample performance of the method. Real data analysis on modelling and classifying disease sub-types demonstrates the applicability of our method as well. 
	\end{abstract}
	
	\noindent%
	{\it Keywords:} Convex optimization, clustering, model selection, dimension reduction.
	\vfill
	
	\newpage
	\spacingset{1.9} 
	\section{Introduction}
	\setcounter{equation}{0}
	Let $\{X_t\}$ be a categorical time series in discrete time, with finite state space $\Sigma$. We suppose that the evolution of the 
	time series follows an $m$-th order Markov structure, where 
	\beq
	\cL\big(X_{t+1}\big|X_s, s\leq t\big) = \cL\big(X_{t+1}\big| X_s, t-m<s\leq t\big)
	\label{MC-m}
	\eeq
	for some $m\geq 1$.
	Here for any random vectors $X,Y$ defined on a common probability space, we write 
	$\cL(Y|X)$ to denote the probability distribution of $Y$ given $X$. Even when the alphabet
	$\Si$ is small, such as $\Si=\{0,1\}$ in applications involving binary chains or $\Si=\{A,G,T,C\}$
	in genetics applications,  
	complexity of the 
	model \eqref{MC-m} increases fairly quickly and parameter 
	estimation may be difficult 
	even for moderately large $m$. Indeed,
	in the absence of a parametric model specification, 
	the number of free parameters associated with \eqref{MC-m} is given by 
	$|\Si|^m(|\Si|-1),$ which grows geometrically fast in the order $m$, where 
	$|\Si|$ denotes the size of the alphabet, that is the number of elements in $\Si$.
	
	Different dimension reduction strategies have been applied to reduce the model complexity in \eqref{MC-m}, such as Variable Length Markov Chains (VLMC) based on tree-structured conditioning sets. This idea was first introduced by \cite{rissanen1983universal}, where relevant contexts (recent pasts) of variable orders are determined to form a context tree. In VLMC, $P(X_{t+1}=x_{t+1}|X_{t}=x_{t},\ldots,X_1=x_1)=P\big(X_{t+1}=x_{t+1}\big|\Tilde{X}_t^{(\ell)}=\Tilde{x}_t^{(\ell)})$, where $\Tilde{X}_t^{(\ell)}=(X_t,X_{t-1},\ldots,X_{t-\ell+1})$, $\Tilde{x}_t^{(\ell)}$ is the observed value of $\Tilde{X}_t^{(\ell)}$ and the tree length
	$\ell$ may not be a fixed number, but rather is a function of the past values $(x_{t},\ldots,x_1)$. In general, context tree models have $L(|\Si|-1)$ parameters, where $L$ is the number of leaves in the context tree. That $L  \leq |\Si|^{m}$ can take on arbitrary positive integer values for general context trees highlights the flexibility of a model with variable length contexts, and the fact that
	such models can lead to huge reductions in the number of parameters, especially when there is a long context in a single direction. A model of a variable order allows for a better trade-off between bias that arises through using contexts that are too short, and variance that increases with having many parameters, thus improving statistical inference. \cite{buhlmann1999variable} and \cite{buhlmann2000model} developed model selection strategies and studied asymptotic behaviour of Variable Length Markov Chains (VLMC). Recently, \cite{kontoyiannis2020bayesian} and \cite{papageorgiou2022posterior} have developed inference and posterior representations for Bayesian Context Trees (BCT) for discrete time series analysis. These two papers also illustrate prediction in the BCT set-up using a posterior predictive distribution.
	
	\cite{roos2009sparse} and \cite{roos2009estimating} pointed out that there can be relevant contexts that do not have the hierarchical structure of a context tree. Although they have discussed the possibility of more general models, the analyses of those papers were limited to the case where $\Sigma=\{0,1\}$. Recently, researchers began studying sparse models posed in terms of a general partition of the set of all 
	$m$-tuples $\Si^m$, where $m$ is the maximal order of Markovian dependence. Such models are called Sparse Markov Models (SMM), and introduce a sparse parametrization 
	based on an unknown
	grouping of all possible $m$th order histories $\Si^m$. This generalization was first proposed by \cite{garcia2011minimal}, who called it Minimal Markov Models. Later on, \cite{jaaskinen2014sparse} developed Bayesian predictive methods to analyze sequence data using SMMs. \cite{xiong2016recursive}  extended the previous paper, introducing a 
	recursive algorithm for optimizing the partition for an SMM. Following a similar approach, \cite{bennett2023fitting} developed a method for fitting sparse Markov models using a collapsed Gibbs sampler. In this paper, we also consider SMMs in full generality, allowing an arbitrary and unknown number of  groups. Specifically, let 
	$\C_1,\ldots,\C_{k_0}$ be a partition of $\Si^m$. Then, the Markov chain $\{X_t\}$ 
	in \eqref{MC-m} is an SMM with groups $\{\C_1,\ldots,\C_{k_0}\}$ if it 
	satisfies the following  sparse representation:
	\beq
	P\big(X_{t+1}\in \cdot \big | X_t=a_{-1},\ldots,X_{t-m+1}=a_{-m}\big)~\mbox{is the same for all} ~(a_{-m},\ldots,a_{-1})\in \C_i,
	\label{smm}
	\eeq
	for each $i=1,\ldots, k_0$. Thus, for each $i$, the transition probability 
	remains unchanged over all $m$-step histories lying in the set $\C_i$. This 
	reduces the number of unknown probability parameters to $k_0(|\Si|-1)$.
	However, both the number $k_0$ of the sets in the partition and 
	the sets $\C_i$ themselves are unknown and must be estimated from the data. 
	
\begin{figure}[!htbp]
		\centering
		\includegraphics[width=\textwidth]{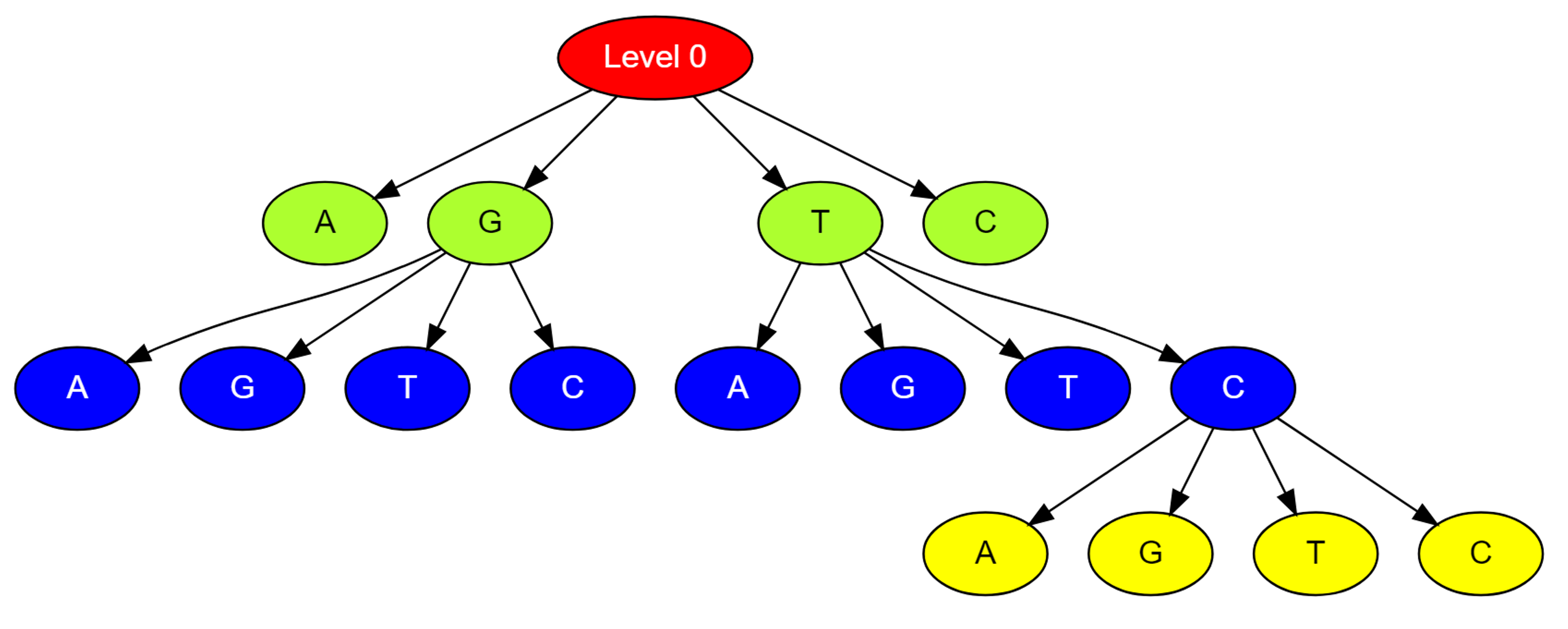}
		\caption{Context Tree for a VLMC of Order $3$}
		\label{fig:vlmc}
	\end{figure}

\begin{figure}[!htbp]
		\centering
		\includegraphics[width=\textwidth]{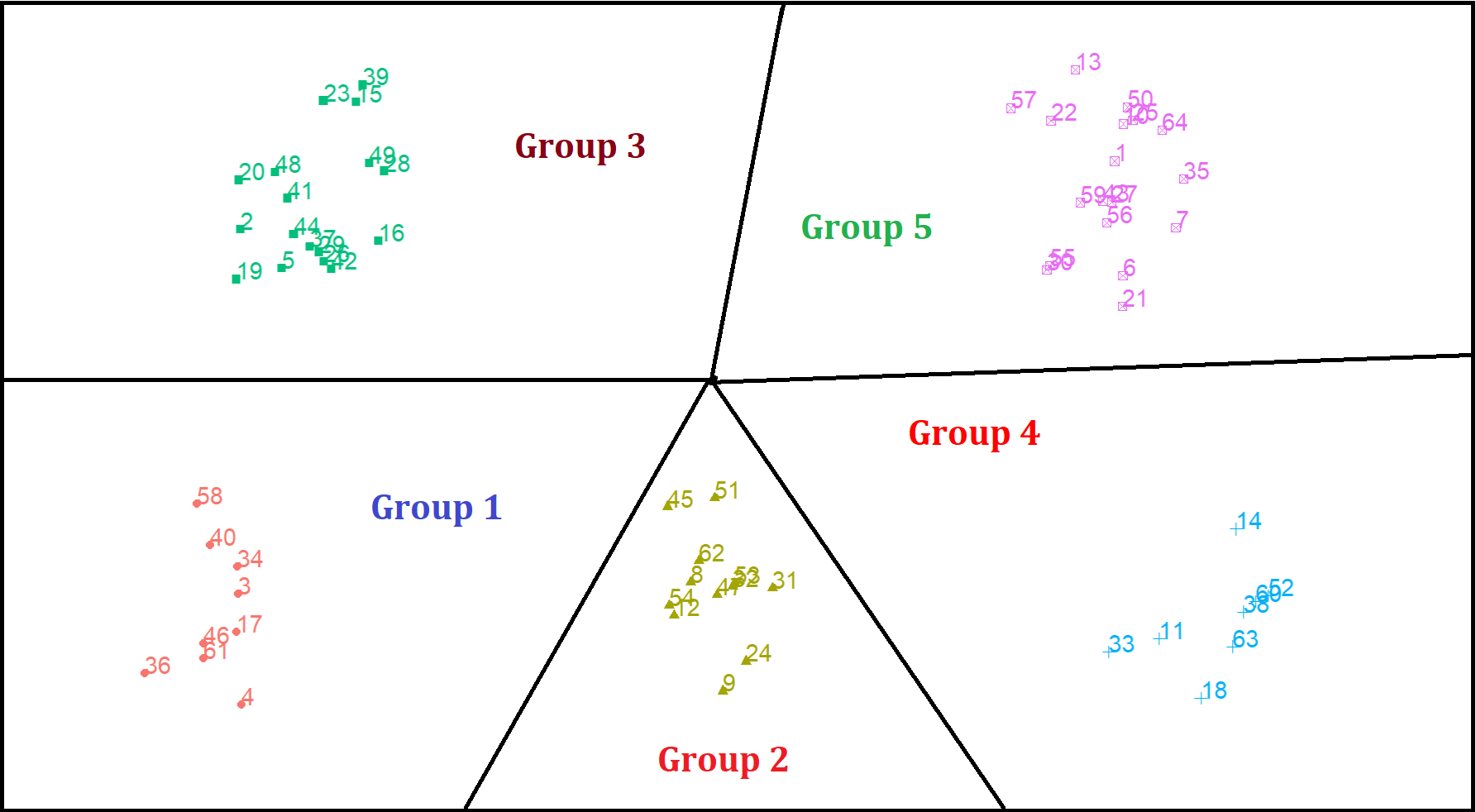}
		\caption{Partition of Triplets for SMM of Order $3$}
		\label{fig:smm}
	\end{figure}

	To illustrate VLMC and SMM, we provide a very simple example of both models using DNA sequences with $\Sigma= \{A, G, T, C\}$ . In figure (\ref{fig:vlmc}), we present the context tree of a VLMC of order $m=3,$ with Level 0 representing the current time $t.$ The tree structure indicates that for all $16$ histories of order $3$ with the most recent history being $x_{t-1}=A$, the transition probability matrices are the same. A similar structure 
	holds true for $x_{t-1}=C.$ If the two recent histories are $x_{t-1}=G$ and $x_{t-2}=A,$ then the transition probability matrices for all $4$ possible triplets $(x_{t-3},A,G)$ are the same; and so on. Hence the given  context tree corresponds to a partition of the 64 3-tuples of $\Sigma^3$ into 12 different groups, with contexts represented by the leaf nodes of the graph. However, for a SMM, the grouping can be arbitrary and does not necessarily have to follow a tree structure. One such example is portrayed in figure (\ref{fig:smm}), where we enumerated the histories of a third-order Markov model as $1,2,3,\ldots,64$. These histories are partitioned arbitrarily into $5$ groups without any tree-like structure, where all histories
	in a given group have the same transition probabilities. Thus, VLMC form a special subclass of SMM.

	
	The generalization to SMM introduces additional 
	challenges for model fitting. Indeed, 
	the task of identifying the true partition is a 
	difficult problem even for moderately large $m$. To appreciate why, note that the total number of partitions of $\Sigma^m$, given by the
	well-known Bell number
	$B(|\Sigma|^m)$, grows at a very fast rate with the order $m$ (cf. \cite{de1981asymptotic}):
	%
	%
	$$
	B\big(|\Sigma|^m\big) \approx \exp \Big( m |\Sigma|^m \log |\Sigma|\Big).
	$$
	For example, with $\Sigma=\{0,1\}$,
	$B(|\Sigma|^3)= 4140$,
	while 
	%
	$B(|\Sigma|^4) = 
	10480142147$.
	As a result, selecting the true partition 
	from such a large collection of partitions is very 
	difficult. 
    
      In this paper, we propose a novel approach for fitting Sparse Markov Models (SMMs) by leveraging recent advances in \emph{convex clustering}. Our method addresses the challenges of determining the true model order and uncovering the latent partition structure in the space of observed $m$-tuples. By treating the \emph{empirical transition probability vectors} as data points, we apply convex clustering techniques---particularly the framework of \cite{chi2015splitting}---to group together those vectors that arise from the same underlying Markovian regime. The core idea is to minimize a penalized criterion function that encourages fusion of similar transition probabilities, leading to automatic partition recovery. The use of an appropriate convex distance metric ensures that the optimization remains tractable even for large values of $m$, and enables scalable implementation. A data-driven BIC-based procedure is used to select the regularization parameter, and we establish \emph{theoretical guarantees} for consistent model recovery. Through a comprehensive simulation study, we demonstrate the strong empirical performance of our method compared to existing approaches for SMM estimation. Finally, we apply our approach to a real-world classification task involving partial DNA sequences from respiratory viruses, achieving significantly lower mis-classification rates than alternative methods.

The rest of the paper is organised as follows. In section \ref{sec_method}, we describe in detail the methodology for fitting SMM using convex clustering approach. Section~\ref{sec_results} presents theoretical results, including model consistency guarantees. Section~\ref{sec_simulation} reports simulation experiments comparing our method with existing techniques. Section~\ref{sec_data analysis} illustrates the application of our method to real virus classification data. Proofs of the theoretical results and some supplementary tables for simulation studies are given in the appendix.

	\section{Methodology} \label{sec_method}
	
	\subsection{Notation} 
	Let 
	$\bbn=\{1,2,\ldots\}$ be the set of all positive integers,
	$\cxn=(X_1,\ldots,X_n)$, and  $\Tilde{X}_t^{(m)}=(X_{t},X_{t-1},\ldots,X_{t-m+1})$, for $m\geq 1$, $t\in \bbn$. Write
	$w$ for an ordered (finite) sequence of $\Si$-elements of length $|w|$.
	Let $w u$ denote the (ordered) concatenation of $w$ and $u$. Write $|\Si|=d$ and w.l.o.g., set $\Sigma=\{1,\ldots,d\}$. Let 
	$\Si^m =\{\si_1,\ldots,\si_p\}$ so that 
	$p=|\Si|^m$. Let $N_w=\sum_{t=|w|}^{n-1} \ind(\Tilde{X}_t^{(|w|)}= w)$
	where $\ind(\cdot)$ denotes the indicator function. For any $S\subset\Sigma^m$ and $a\in \Sigma$, define $N_{S}=\sum_{t=m}^{n-1} \ind(\Tilde{X}_t^{(m)}\in S)$ and $N_{S,a}=\sum_{t=m}^{n-1} \ind(\Tilde{X}_t^{(m)}\in S,X_{t+1}=a)$. In particular, $N_{\sigma_j}$ denotes the number of times the chain $\Tilde{X}_t^{(m)}$ hits the $m$-tuple $\sigma_j$, and $N_{\sigma_j,a}$ is the number of transitions from $\sigma_j$ to $a$. Note that $n-m+1$ denotes the total number of $m$-th order histories in the observed variables $\cxn$. 
	
	Next we define the probabilities associated with the SMM \eqref{smm}. For 
	$j=1,\ldots,p$ and $a\in \Si$, let 
	\beas
	\pi_{j,a} = P\big(X_{t+1}=a\big|\Tilde{X}_t^{(m)}=\si_j\big);
	\eeas
	and $\boldsymbol\pi_{j} = (\pi_{j,a})_{a=1,\ldots,d}$ let be the corresponding transition probability vector. Note that by the SMM property, for any $a\in \Si$, 
	the transition probability $\pi_{j,a}$ is a constant over all $j$ such that $\si_j\in \C_i$. However, we do not know the sets $\C_i$ and determining them is one of the challenges of fitting an SMM to a data set. To that end, define non-parametric estimators of $\pi_{j,a}$ using their empirical versions:
	$$
	\hat{\pi}_{j,a}= N_{\si_j,a}/N_{\si_j},
	$$
	and let $\hat{\boldsymbol\pi}_j$ be  the transition probability vectors consisting of the elements $\hat{\pi}_{j,a}$.
	
	Here we propose a new approach to fitting the SMM  based on regularization. 
	\subsection{Description of the Method}
	Consider the penalized criterion function
	\beq
	\dfrac{1}{2}\sum_{j=1}^p \lVert \hat{\boldsymbol\pi}_j - \mathbf{b}_{j} \rVert_2^2
	+\la\sum_{1\leq i<j\leq p} w_{i,j} \rho(\mathbf{b}_{i}, \mathbf{b}_{j}) 
	\label{cr1}
	\eeq
	over $\mathbf{b}_{j}=(b_{j,1},\ldots, b_{j,d})^T \in \Pi_d$ for $j=1,\ldots, p$, 
	where $\la>0$ is a penalty parameter, $w_{i,j}$ are suitable nonnegative weights, $\Pi_d$ is the $d$-dimensional simplex $\Pi_d=\{(u_1,\ldots,u_d)\in [0,1]^d :
	u_1+\ldots+u_d=1\}$ and where $\rho(\cdot,\cdot)$ is a distance measure 
	between two $d$-dimensional probability vectors. 
	Thus, \eqref{cr1} treats the estimators $\hat{\pi}_{j,a}$ as  (correlated) ``observations''
	and 
	penalizes the distance between all distinct pairs of probability vectors in order to identify
	the identical probability vectors. In particular, the number of parameters 
	grows at a rate proportional to the size of the true
	partition in the SMM  and with a suitable choice of the 
	penalization term, one can identify the identical probability vectors.
	When  $\rho(\mathbf{b}_{i}, \mathbf{b}_{j})^2=\sum_{a=1}^d (b_{i,a} -
	b_{j,a})^2$, \eqref{cr1} gives a version of the Group LASSO of \cite{yuan2006model}
	that is designed for selecting   pairs of full vectors   that are close, and we have a convex 
	optimization problem that can be solved for large $p.$  On the other hand,
	if we use  the $\ell_1$ distance $\rho(\mathbf{b}_{i}, \mathbf{b}_{j})=\sum_{a=1}^d |b_{i,a} -
	b_{j,a}|$, then only component-wise zero differences can be identified. 

	Once we minimize the criterion function in (\ref{cr1}), it is a relatively easy task to 
	find estimates of $k_0$ and the sets $\C_i$. Specifically, we start with a pair with the smallest $i$
	and seek all $j>i$ such that the distance between the solutions $\mathbf{b}^{*}_{i}$ and $\mathbf{b}^{*}_{j}$ 
	is zero. Then, we set $\hat{\C}_1$ to be the set consisting of $i$ and all such $j$. In the next step,
	we consider all pairs that are not in $\hat{\C}_1$ and repeat the procedure until all pairs with 
	estimated zero distances have been grouped. In case there are indices $j$ for which 
	none of the estimated paired distances are zero, we keep them as  singletons, that is groups 
	consisting of   single elements. This gives the estimated groups $\hat{\C}_i : i=1,\ldots, \hk$,
	with $\hk$ giving an estimate of $k_0$. 
	
	In comparison, traditional clustering methodologies like $K$-means have many limitations. In most cases, we have to pre-specify the number of clusters, along with the possibility that we end up with a local minima instead of the global one. The advantage of clustering  by solving equation (\ref{cr1}) for a range of $\lambda$ is that we get a solution path from at most $p$ many singleton clusters to only one cluster consisting of all the elements. Subsequently, we can fix some criterion function which will enable us to find the optimum cluster assignment among all possible models in the solution path. Hence, not only do we not need to fix the number of clusters beforehand, but we also avoid the problem of being stuck at local minima. This particular approach will be  broadly 
	referred to as ``{Convex Clustering}".
	
	\subsection{Computational considerations
		%
	}
	Several efficient algorithms have been developed in recent years to solve equation (\ref{cr1}) when the penalty function $\rho$ is convex; e.g. $\rho(\mathbf{b}_{i},\mathbf{b}_{j})=\lVert \mathbf{b}_{i}-\mathbf{b}_{j} \rVert_p$ for some $p\ge 1$. 
	\cite{pelckmans2005convex}, \cite{lindsten2011just}, \cite{hocking2011clusterpath} and others recently proposed this convex clustering approach and established it to be more robust and scalable in comparison to the traditional approaches. \cite{lindsten2011just} used an off-the-shelf convex solver CVX to solve the convex clustering problem, which suffers from scalability issues. Theoretical perfect cluster recovery conditions have been derived by \cite{zhu2014convex} only for two clusters, while \cite{panahi2017clustering} derived perfect recovery conditions for general $k$ clusters, but under the assumption of uniform weights. \cite{sun2021convex} provided sufficient conditions for theoretical recovery conditions under more general weight choices. They have also developed a faster algorithm called semismooth Newton based augmented Lagrangian method (SS-NAL), and derived the convergence criteria for their algorithm. Recently, \cite{wang2021integrative} have introduced 
	the Integrative Generalized Convex Clustering Optimization (iGecco) method for solving the convex clustering problem for more general loss functions, including non-differentiable ones. 
	The major difference of our set-up from previous developments is that we cluster empirical transition probability vectors as opposed to the original data points.
	
While methods like iGecco \citep{wang2021integrative} have been developed for more general loss functions and are particularly useful in integrative clustering across multiple data modalities, our setup is more specialized. Specifically, we focus on clustering empirical transition probability vectors derived from categorical sequences using a squared error loss. The formulation of \cite{chi2015splitting} aligns directly with this objective and provides efficient solvers tailored to squared loss with convex penalties. In contrast, iGecco’s flexibility for general loss functions introduces unnecessary complexity in our context. Therefore, we opt for the simpler and more specialized algorithm of \cite{chi2015splitting}, which has demonstrated strong performance under our model assumptions.

We now introduce the specific computational algorithm proposed by \cite{chi2015splitting} that we use to minimize the objective function in (\ref{cr1}).  For $\rho(\mathbf{x})=\lVert \mathbf{x} \rVert_2$, we first view solving  equation (\ref{cr1}) as the following constrained optimization problem 
	\begin{equation}\label{eq_constrained}
		\begin{aligned}
			\min & \dfrac{1}{2}\sum_{j=1}^{p} \lVert \hat{\boldsymbol\pi}_j-\mathbf{b}_j \rVert_2^2 + \lambda \sum_{l\in\mathcal{E}} w_l \lVert \mathbf{v}_l \rVert_2\\
			\text{subject to }& \mathbf{b}_{l_1}-\mathbf{b}_{l_2}-\mathbf{v}_l=0;
		\end{aligned}
	\end{equation}
	where $\mathcal{E}$ is the set of all distinct edges $\{l:l=(l_1,l_2),l_1<l_2,w_l>0\}$. Here, a new splitting variable $\mathbf{v}_l$ has been introduced to capture the difference between the group centroids, which makes the optimization procedure much easier. Two algorithms have been developed for solving this constrained optimization problem, namely alternating direction method of multipliers (ADMM) and alternating minimization algorithm (AMA). ADMM solves constrained optimization problems by breaking them into smaller subproblems that are easier to solve, alternating between primal and dual updates.
AMA simplifies the process further by avoiding auxiliary variable updates, leading to significantly faster convergence under sparse weights. For both of these algorithms,  first we incorporate an augmented Lagrangian as follows:
	\begin{equation}\label{eq_augment_lang}
		\begin{aligned}
			\mathcal{L}_{\nu}(\mathbf{B},\mathbf{V},\boldsymbol\Gamma) = & \dfrac{1}{2}\sum_{j=1}^{p}  \lVert \hat{\boldsymbol\pi}_j-\mathbf{b}_j \rVert_2^2+\lambda \sum_{l\in\mathcal{E}} w_l \lVert \mathbf{v}_l \rVert_2\\
			& + \sum_{l\in\mathcal{E}} \langle\boldsymbol\gamma_l, \mathbf{v}_l-\mathbf{b}_{l_1}+\mathbf{b}_{l_2} \rangle + \dfrac{\nu}{2}\sum_{l\in\mathcal{E}} \lVert \mathbf{v}_l-\mathbf{b}_{l_1}+\mathbf{b}_{l_2} \rVert_2^2,
		\end{aligned}
	\end{equation}
	where $\mathbf{B},\mathbf{V}$ and $\boldsymbol\Gamma$ are the matrices with $\mathbf{b}_j,\mathbf{v}_l$ and $\boldsymbol\gamma_l$ for $j=1,\ldots,p$ and $l\in\mathcal{E}$ in their columns respectively. Splitting the variables in such fashion would allow us to update $\mathbf{B}$, $\mathbf{V}$ and $\boldsymbol\Gamma$ sequentially, given the other variables. The convergence of ADMM does not depend on the choice of $\nu$; it is known to converge for any $\nu>0$. On the other hand, AMA converges for any $0<\nu<2/p$. 

	Since AMA provides much faster results, we 
	will use this algorithm in numerical implementation of
	our methodology. Suppose, $\mathbf{B}^{(t)}$ and $\boldsymbol\Gamma^{(t)}$ be the parameter values in the $t^{th}$ step. The updates in the next step are computed using the following relations:
	\begin{equation*}
		\begin{aligned}
			\mathbf{b}_j^{(t+1)}&=\hat{\boldsymbol\pi}_j+ \sum_{l_1=j} \boldsymbol\gamma_l^{(t)}- \sum_{l_2=j} \boldsymbol\gamma_l^{(t)}\\
			\boldsymbol\gamma_l^{(t+1)} & = \mathcal{P}_{C_l}(\boldsymbol\gamma_l^{(t)}-\nu \mathbf{g}_l^{(t+1)})
		\end{aligned}
	\end{equation*}
	where $\mathbf{g}_l^{(t+1)}=\mathbf{b}_{l_1}^{(t+1)}-\mathbf{b}_{l_2}^{(t+1)}$, $C_l=\{\boldsymbol\gamma_l:\lVert \boldsymbol\gamma_l \rVert_2\le \lambda w_l\}$, and $\mathcal{P}_A(\mathbf{x})$ is the projection of $\mathbf{x}$ onto the set $A$. We continue until convergence, and the convergence criterion can be formulated using the dual problem and duality gap. 
	\begin{algorithm}
		\caption{AMA}\label{alg:cap}
		Initialize $\boldsymbol\Gamma^{(0)}$
		\begin{algorithmic}[1]
			\For{$t=1,2,3,\ldots$}
			\For{$j=1,2,3,\ldots,p$}
			\State $\boldsymbol\Delta_j^{(t)}=\sum_{l_1=j}\boldsymbol\gamma_l^{(t-1)}-\sum_{l_2=j}\boldsymbol\gamma_l^{(t-1)}$
			\EndFor
			\ForAll{$l$}
			\State $\mathbf{g}_l^{(t)}=\hat{\boldsymbol\pi}_{l_1}-\hat{\boldsymbol\pi}_{l_2} +\boldsymbol\Delta_{l_1}^{(t)}-\boldsymbol\Delta_{l_2}^{(t)}$
			\State $\boldsymbol\gamma_l^{(t)}=\mathcal{P}_{C_l}(\boldsymbol\gamma_l^{(t-1)}-\nu \mathbf{g}_l^{(t)})$
			\EndFor
			\EndFor
		\end{algorithmic}
	\end{algorithm}

	\subsection{Selection of the Tuning Parameter }
	So far, we have discussed the numerical methods to solve (\ref{cr1}) for a given $\lambda$. But it is important to choose an optimum value of $\lambda$ for the optimization problem. In this section, we propose a data driven method to select this tuning parameter using the BIC criterion. For a given $\lambda$, denote the obtained clusters as $\hat{\mathcal{C}}_1(\lambda),\ldots, \hat{\mathcal{C}}_{k_\lambda}(\lambda)$, where $k_\lambda$ is the number of clusters. Define the common transition probability for the $m$-tuples in the estimated group $\hat{\mathcal{C}}_\alpha(\lambda)$ as 
	$$
	\hat{R}^{(\lambda)}_{\alpha,a}=\dfrac{\sum_{\sigma_j\in\hat{\mathcal{C}}_\alpha(\lambda)}N_{\sigma_j,a}}{\sum_{\sigma_j\in\hat{\mathcal{C}}_\alpha(\lambda)}N_{\sigma_j}}=\dfrac{N_{\hat{\mathcal{C}}_\alpha(\lambda),a}}{N_{\hat{\mathcal{C}}_\alpha(\lambda)}}\quad\quad \forall \alpha=1,\ldots,k_{\lambda};a\in \Sigma.
	$$
	The log-likelihood of the observations under the obtained cluster assignment for a particular $\lambda$ is given by
	$$
	\ell_n(\lambda)=\sum_{\alpha=1}^{k_\lambda}\sum_{a\in\Sigma} N_{\hat{\mathcal{C}}_\alpha(\lambda),a}\log \hat{R}^{(\lambda)}_{\alpha,a}.
	$$
	Hence, the BIC score corresponding to the obtained model is 
	$$
	BIC_n(\lambda)=-2\ell_n(\lambda)+k_{\lambda}(|\Sigma|-1)\log n.
	$$
	By a grid search over a range of possible $\lambda$ values, we select the $\lambda$ for which BIC is minimized. The solution of equation (\ref{cr1}) corresponding to that $\lambda$ is considered as the estimated cluster assignment. The novelty of our method is that we are able to select the optimum tuning parameter from the data itself. For general convex clustering scenarios, one may not be able to compute the BIC criterion since the distributional properties of the data points in a cluster are unknown.

The assumption of Markovian structure is useful in our set-up to formulate the likelihood function. Moreover, the CLT-type results provide us the asymptotic distributions of the estimated transition probabilities. In the next section, we provide new theoretical results to demonstrate the model selection consistency under this BIC-based approach. We also provide some theoretical results and conditions under which the true partitions will appear in the solution path of convex clustering for a range of $\lambda$ for large $n$.

	\paragraph{Alternative $\lambda$ Selection via Pathwise Regularization.}
In addition to the BIC-based grid search method, we also explored a pathwise convex clustering approach inspired by the algorithmic regularization framework of \cite{weylandt2020dynamic}. This technique efficiently constructs the solution path for varying values of $\lambda$ without requiring full re-optimization at each point, by leveraging warm starts and algorithmic early stopping. Since this method produces the full solution path in the form of a dendrogram, we continue to use the BIC criterion to select the optimal model within this framework. In our experiments (see Section~\ref{sec_simulation}), the pathwise method yielded slightly lower clustering accuracy compared to the original BIC-guided convex clustering, but achieved substantial computational speed-ups. These properties make the pathwise strategy a promising alternative for high-dimensional categorical time series. While we primarily rely on the BIC criterion for theoretical tractability, the pathwise algorithm provides a scalable and empirically viable option.

	\section{Conditions and Theoretical Results}\label{sec_results}
	\subsection{Conditions}
	
	We consider equation (\ref{cr1}) with $\rho(\mathbf{b}_{i},\mathbf{b}_{j})=\lVert \mathbf{b}_{i}-\mathbf{b}_{j} \rVert_2$. Let  the optimum solution be  denoted by $\mathbf{b}_{i}^{*}(\lambda)$, for $i=1,2,\ldots,p$. Also, let the true partition of the state space $\Sigma^m$ be  $\{\C_1,\ldots, \C_{k_0}\}$,  with the corresponding transition probability vectors being $\bm{R}_1,\ldots,\bm{R}_{k_0}$. Thus,  $\bm{R}_{\alpha,a}=P(X_{t+1}=a\big|Y_t=\sigma_{\alpha})$. Set  $p_{\alpha}=\big|\C_{\alpha}\big|$, the size of the $\alpha^{th}$ partition. Following the notation of \cite{sun2021convex}, define
	\beas
	&& w_i^{(\beta)}= \sum_{j\in \C_{\beta} } w_{i,j} \quad \forall i=1,2,\ldots,p; \quad\quad \mu_{i,j}^{(\alpha)}=\sum_{\ell\ne\alpha} \big\lvert w_i^{(\ell)}-w_j^{(\ell)}  \big\rvert \quad \quad \forall \alpha=1,2,\ldots,k_0;\\
	&& w^{(\alpha,\beta)}=\sum_{i\in \C_{\alpha} }\sum_{j\in \C_{\beta} } w_{i,j} \quad \forall \alpha \ne \beta , \alpha,\beta \in\{1,2,\ldots,k_0\};\quad\quad \hat{\bar{\boldsymbol\pi}}^{(\alpha)}=\dfrac{1}{p_\alpha}\sum_{i\in\C_{\alpha}}\hat{\boldsymbol\pi}_i;\\
	&& \lambda_{\text{min}}^{(n)}= \max_{1\le\alpha\le k_0} \max_{i,j\in\C_{\alpha}}\Bigg\{\dfrac{\lVert \hat{\boldsymbol\pi}_i-\hat{\boldsymbol\pi}_j\rVert_2}{p_\alpha w_{i,j}-\mu_{i,j}^{(\alpha)}}\Bigg\};\\
	&& \lambda_{\text{max}}^{(n)}= \min_{1\le\alpha<\beta\le k_0} \Bigg\{\dfrac{\lVert \hat{\bar{\boldsymbol\pi}}^{(\alpha)}-\hat{\bar{\boldsymbol\pi}}^{(\beta)}\rVert_2}{\frac{1}{p_{\alpha}}\sum_{l\ne\alpha}w^{(\alpha,l)}+\frac{1}{p_{\beta}}\sum_{l\ne\beta}w^{(\beta,l)}}\Bigg\}.
	\eeas
	We shall suppose
	that the following conditions hold.
	\\[.1 in]
	\noi
	{(A1)}~
	$w_{i,j}= w_{j,i}$ and $w_{i,j}>0$ for any $i,j\in \C_{\ell}$, $\ell=1,2,\ldots,k_0$.\\
	[.1 in]
	\noi
	{(A2)}~ $p_\alpha w_{i,j}>\mu_{i,j}^{(\alpha)}$, $\forall i,j\in \C_{\alpha}$ and $\forall \alpha=1,2,\ldots,k_0$. 
	\\[.1 in]
	In (A1) we assume symmetry, and that the weight is positive between two $m$-tuples belonging to the same partition. (A2) gives a lower bound for the weight between two $m$-tuple in a particular group.
	Similar  conditions have been used 
	by \cite{sun2021convex} to prove 
	perfect recovery
	results. In Theorem \ref{thm_special_case} below, 
	we provide some simple sufficient conditions on weight choices to satisfy these conditions. 
	

	Before going into the main results, we state two auxiliary results that will be used for the subsequent 
	results. 
	
	\begin{proposition} \label{prop_1}
		Let $\{X_n\}_{n\geq 1}$ be an aperiodic and irreducible SMM of order $m$ with true  partition $\{\mathcal{C}_1,\ldots, \mathcal{C}_{k_0}\}$. Then, as $n\to\infty$,
		\begin{itemize}
			\item [(a)]  $\hat{\boldsymbol\pi}_j\xrightarrow[]{p}\bm{R}_{\alpha}$~ for ~$j\in \mathcal{C}_\alpha$;
			\item[(b)] $\dfrac{N_{\sigma_j}}{N}\xrightarrow[]{p} q_j$, where $q_j$ is the stationary probability of the state $\sigma_j$;
			\item[(c)] 
			With $\Sigma_\alpha=diag(\bm{R}_\alpha)-\bm{R}_\alpha\bm{R}_\alpha^{(T)}$,
			$$
			\sqrt{N_{\sigma_j}} (\hat{\boldsymbol\pi}_j-\bm{R}_{\alpha})\xrightarrow[]{d} \mathcal{N}(\bm{0},\Sigma_{\alpha}).
			$$
			Since $\Sigma_\alpha$ is of rank $|\Sigma|-1$, the asymptotic Normal distribution is singular.
		\end{itemize}
	\end{proposition}
	Thus, Proposition (\ref{prop_1}) asserts  weak consistency and asymptotic normality of  the estimated transition probability vectors, which can be proved using existing results on Markov chains in \cite{billingsley1961statistical}. 
	The next result deals with perfect recovery  under general weight choices under Conditions (A.1) and (A.2).

	\begin{proposition}\label{prop_2}
		Suppose the above conditions (A1) and (A2) hold and $\lambda_{\text{min}}^{(n)} < \lambda_{\text{max}}^{(n)}$. Then for any $\lambda\in (\lambda_{\text{min}}^{(n)}, \lambda_{\text{max}}^{(n)})$, $\mathbf{b}_i^{*}(\lambda)=\mathbf{b}_j^{*}(\lambda)$ for $i,j\in \mathcal{C}_{\alpha}$; $\alpha=1,..,k_0$ and $\mathbf{b}_i^{*}(\lambda)\ne\mathbf{b}_j^{*}(\lambda)$ for any $i\in \mathcal{C}_{\alpha},j\in \mathcal{C}_{\beta},\alpha\ne\beta$. In other words, for any $\lambda\in (\lambda_{\text{min}}^{(n)} , \lambda_{\text{max}}^{(n)})$, we recover the true partition of the state space.
	\end{proposition}  
	These propositions will be among the key tools used for proving our results. The CLT result will be useful for determining  probability bounds for perfect recovery under the conditions of the proposition (\ref{prop_2}). In the next subsection, we state our major theoretical findings.
\subsection{Main Results}
	
	Although the solution to the objective function in equation~(\ref{eq_constrained}) involves optimization over vectors in $\mathbb{R}^d$, the feasible set is restricted to the probability simplex $\Pi_d$. As a result, the optimal solutions $\mathbf{b}_j^*(\lambda)$ remain valid probability distributions over $\Sigma$—that is, nonnegative vectors summing to one. We formalize this below and provide a short proof in the appendix.

\begin{lemma} \label{thm_prob_dist}
For any $\lambda > 0$, the optimal solution $\mathbf{b}_i^*(\lambda)$ lies in the $d$-dimensional probability simplex $\Pi_d$, i.e.,
\begin{itemize}
    \item[(a)] $b^*_{i,a}(\lambda) \ge 0$ for all $a = 1,\ldots,d$, and
    \item[(b)] $\sum_{a=1}^d b^*_{i,a}(\lambda) = 1$.
\end{itemize}
\end{lemma}

This result follows directly from standard properties of convex optimization over closed convex sets (see \cite{boyd2004convex}) and the structure of the algorithm used (see \cite{chi2015splitting}). While one could impose simplex constraints explicitly during optimization, our augmented Lagrangian approach already preserves this structure without additional computational cost.

Despite this, we do not directly use the centroids $\mathbf{b}_j^*(\lambda)$ from convex clustering as estimates of the transition probabilities. Due to the shrinkage effect induced by the fusion penalty, these cluster centers may be biased toward the global average and not accurately represent the true transition distributions. Therefore, once the partition is obtained, we re-estimate the transition probabilities for each cluster using the empirical frequencies of the histories in that cluster. This approach improves estimation accuracy while preserving the benefits of regularized clustering.
	
	Next, we would like to derive the probability of true cluster recovery. There are two steps involved in this process. First, we need the true model in the solution path over varying $\lambda$. This implies the conditions of Proposition \ref{prop_2} must be satisfied, i.e. $\lambda_{\text{min}}^{(n)} < \lambda_{\text{max}}^{(n)}$. From Theorem \ref{thm_cluster_recovery}, which is given next, we get a lower bound on the probability of the true model being present in the solution path. Note that \cite{sun2021convex} have derived these perfect recovery conditions for a given fixed data set when the data points in a particular cluster are close to each other. Our approach is significantly different, as we cluster estimated transition probability vectors, which are random variables, and thus $\lambda_{\text{min}}^{(n)}$ and $\lambda_{\text{max}}^{(n)}$ are random variables as well. We provide theoretical bounds to ensure that the probability of the event $\big\{\lambda_{\text{min}}^{(n)}<\lambda_{\text{max}}^{(n)}\big\}$ is $1-\mathcal{O}_p(e^{-n})$.
	
	\begin{theorem}\label{thm_cluster_recovery}
		Define 
		$$
		\begin{aligned}
			\delta&=\min_{1\le\alpha<\beta\le k_0}\lVert \bm{R}_\alpha-\bm{R}_\beta\rVert_2;\quad \quad
			\delta_1=\min_{1\le\alpha\le k_0}  \min_{i,j\in\mathcal{C}_{\alpha}}\big(p_\alpha w_{i,j}-\mu_{i,j}^{(\alpha)}\big);\\
			\delta_2&=\max_{1\le\alpha<\beta\le k_0} \Big( \frac{1}{p_{\alpha}}\sum_{l\ne\alpha}w^{(\alpha,l)}+\frac{1}{p_{\beta}}\sum_{l\ne\beta}w^{(\beta,l)}\Big).
		\end{aligned}
		$$
		Then, under Conditions (A1) and (A2), as $n\to\infty$,
		$$
		\begin{aligned}
			P\Big(\lambda_{\text{min}}^{(n)} < \lambda_{\text{max}}^{(n)} \Big) &\ge P\Big(\lVert\hat{\boldsymbol\pi}_j-\bm{R}_\alpha\rVert_2<\dfrac{\epsilon}{2} \forall j\in\mathcal{C}_\alpha,\forall\alpha=1,\ldots,k_0\Big)\\
			&\ge 1-\sum_{\alpha=1}^{k_0}C_{1}^{(\alpha)}\sum_{j\in\mathcal{C}_\alpha}\exp\Big[-(n-m)\epsilon^2C_{2,j}\Big]
		\end{aligned}
		$$
		for $0<\epsilon<\dfrac{\delta\delta_1}{\delta_1+\delta_2}$, and for some constants $C_1^{(\alpha)},C_{2,j}>0$.
	\end{theorem}
	
	Looking at the expressions for $\lambda_{\text{min}}^{(n)}$ and $\lambda_{\text{max}}^{(n)}$, it is evident that $\lambda_{\text{min}}^{(n)}$ shrinks towards $0$ as $n$ increases as the estimated transition probability vectors $\hat{\boldsymbol\pi}_i$ and $\hat{\boldsymbol\pi}_j$ belonging to the same cluster $\C_\alpha$ become closer to each other. On the other hand, the different group means $\hat{\bar{\boldsymbol\pi}}^{(\alpha)}$ and $\hat{\bar{\boldsymbol\pi}}^{(\beta)}$ tend to get separated from each other, making $\lambda_{\text{max}}^{(n)}$ converge to a positive number, so that eventually we get $\lambda_{\text{min}}^{(n)}<\lambda_{\text{max}}^{(n)}$. These expressions also tell us that in order to have perfect recovery of the clusters, a scaled version of the maximum within-group deviation of the transition probabilities should be less than a scaled version of the minimum between-group variation. These scales are heavily dependent on the choice of the weights $w_{i,j}$. Note that, if we choose the weights in a way so that $w_{i,j}$ is higher if $\hat{\boldsymbol\pi}_i$ and $\hat{\boldsymbol\pi}_j$ are closer (and potentially belong to the same cluster), and lower if they are far from each other (potentially belonging  to different clusters), the denominator of the term $\lambda_{\text{min}}^{(n)}$ will be higher, and the denominator of $\lambda_{\text{max}}^{(n)}$ will be lower in the ideal scenario. Hence, this particular choice of the weights will enhance separating $\lambda_{\text{min}}^{(n)}$ and $\lambda_{\text{max}}^{(n)}$, increasing the chance of recovering the true cluster assignment. Although our theoretical guarantees are stated over a continuous subset of the real line for $\lambda$, in actual implementation we operate over a discrete grid of $\lambda$ values. This is because the theoretical range depends on unknown model parameters and serves only as a sufficient condition for recovery. 
    
    Once we have the true model in the solution path, the next step is to establish that the probability of selecting that model through the BIC criterion converges to $1$ as $n\to\infty$. The next theorem 
	gives a precise statement of this result.
	
	\begin{theorem} \label{thm_BIC}
		Suppose the conditions of Theorem (\ref{thm_cluster_recovery}) hold, and $\lambda_{\text{min}}^{(n)} < \lambda_{\text{max}}^{(n)}$. For any $\lambda$, denote the clustering assignment obtained by minimizing the equation (\ref{cr1}) as $M_{\lambda}=\{\hat{\C}_1(\lambda),\ldots, \hat{\C}_{k_\lambda}(\lambda)\}$; where $k_\lambda$ is the associated number of partitions of the $m$-tuples. Suppose $\ell_n(\lambda)$ is the log-likelihood of the observations corresponding to cluster assignment $M_{\lambda}$, and the corresponding BIC score is $BIC_n(\lambda)=-2\ell_n(\lambda)+k_{\lambda}(d-1)\log n$. Choose some $\lambda_0\in (\lambda_{\text{min}}^{(n)} , \lambda_{\text{max}}^{(n)})$. Then, for any $\lambda$ such that $M_{\lambda}\ne M_{\lambda_0}$,
		$$
		P\Big(BIC_n(\lambda_0)<BIC_n(\lambda)\Big)\longrightarrow 1
		$$
		as $n\to \infty$.
	\end{theorem}
	
	Theorem \ref{thm_BIC} asserts the consistency of the model selection using the BIC criterion and is one of the most important conclusions of this paper. Variable selection consistency results in \cite{zhang2010regularization} use a similar BIC-criterion for the LASSO penalty in multiple linear regression. However, our set-up is very different from the regression set-up (with independent observations).
	Further, the penalty function is quite different.
	As a result, the key steps for proving our result are very different. The consequence of this theoretical result is extremely important in applications. Even for moderately large sample sizes, we can achieve good clustering performance. We will demonstrate these properties for finite samples in the simulation study. 
	
	Although we have stated our results under conditions (A.1) and (A.2), we still need to check whether these conditions are feasible in practice. The next result provides sufficient conditions for perfect cluster recovery under a particular weight choice involving Gaussian kernels that also 
	produces good clustering results in finite samples. 
	
	\begin{theorem} \label{thm_special_case}
		Define $p_{min}=\min_{\alpha}p_{\alpha},p_{max}=\max_{\alpha}p_{\alpha}$, and assume that the true cluster or partition sizes are different for the SMM. Suppose that
		\begin{itemize}
			\item [(a)]
			$w_{i,j}=e^{-\phi\lVert \hat{\boldsymbol\pi}_i-\hat{\boldsymbol\pi}_j\rVert_2^2}l^k_{i,j}$, where $l^k_{i,j}$ is the indicator function that $\hat{\boldsymbol\pi}_i$ is one of the $k$ nearest neighbours of $\hat{\boldsymbol\pi}_j$ or vice versa, for some $\phi>0$;
			\item [(b)] $k\ge p_{max}-1$;
			\item [(c)] for some $\epsilon<\epsilon_{max}=\dfrac{\delta}{2}-\dfrac{1}{2\phi\delta}\log\Big(2\Big(\dfrac{k'+1}{p_{min}}-1\Big)\Big)$,
			$\lVert\hat{\boldsymbol\pi}_j-\bm{R}_\alpha\rVert_2<\dfrac{\epsilon}{2};$ $\forall j\in\mathcal{C}_\alpha$, $\forall\alpha=1,\ldots,k_0$, where 
			$$
			k'=\max_{i} \sum_{j=1}^{p}l^k_{i,j}.
			$$
		\end{itemize}
		Then conditions (A1) and (A2) are satisfied. Moreover, $\delta_1\ge p_{min}e^{-\phi\epsilon_{max}^2}-2(k'+1-p_{min})e^{-\phi(\delta-\epsilon_{max})^2}=\delta^{(min)}_{1}$ and $\delta_2\le 2(k'+1-p_{min})e^{-\phi(\delta-\epsilon_{max})^2}=\delta_2^{(max)}.$
	\end{theorem}
	Theorem \ref{thm_special_case} simplifies  Conditions (A1) and (A2) for a special choice of weights, which we will use later in our simulation studies. The intuition behind this  choice is that $w_{i,j}$ should be a decreasing function of $\lVert\hat{\boldsymbol\pi}_i-\hat{\boldsymbol\pi}_j\rVert_2,$ which enforces less penalization for well-separated points. The following results
	are direct consequences of  Theorem \ref{thm_special_case}:
	\begin{corollary} \label{cor_1}
		Under the assumptions of Theorem \ref{thm_cluster_recovery},
		\begin{itemize}
			\item [(a)] $\lambda_{\text{min}}^{(n)}\le \dfrac{\epsilon}{\delta^{(min)}_{1}}$, $\lambda_{\text{max}}^{(n)}\ge \dfrac{\delta-\epsilon}{\delta_2^{(max)}}$.
			\item[(b)] $\epsilon<\min\Big\{\epsilon_{max},\dfrac{\delta\delta_1}{\delta_1+\delta_2}\Big\}\implies\lambda_{\text{min}}^{(n)}<\lambda_{\text{max}}^{(n)}$.
		\end{itemize}
	\end{corollary}
	
	\begin{corollary}\label{cor_2}
		For a balanced design, i.e. when $p_{\alpha}=p/k_0$ are the same for all groups $\mathcal{C}_\alpha$, $\delta_2^{(max)}=0$ if $k=p/k_0-1$. Hence, for any $\epsilon<\dfrac{\delta}{2}$, perfect recovery is possible for $\lambda\in (\lambda_{\text{min}}^{(n)},\infty)$.
	\end{corollary} 
	We present all proofs in the appendix section. Corollary \ref{cor_1} gives us an idea of
	how close the empirical transition probabilities for each $m$-tuple are  to the true probability
	vectors. Corollary \ref{cor_2} considers a special case when the design is balanced. In that scenario, for large $n$ and for the correct choice of the nearest neighbour, the true model can be retrieved for a wide range of tuning parameters $\lambda$, thereby 
	providing very accurate 
	clustering results. In the next section,
	we will explore the impact of weight choices on clustering accuracy through simulations of finite samples.
	
	\section{Simulation Study}\label{sec_simulation}

In this section, we numerically evaluate the performance of the convex clustering methodology described earlier, focusing on its ability to recover the true cluster assignments. We compare clustering results across different choices of the weights $w_{i,j}$, considering sparse Markov models (SMMs) of varying orders, sequence lengths, and alphabet sizes $|\Sigma|$. Additionally, we benchmark our approach against several competing methods to demonstrate its practical utility. Specifically, we compare convex clustering with the BIC-based hierarchical clustering proposed by \cite{garcia2011minimal} (GGL), the Bayesian factor hierarchical clustering by \cite{xiong2016recursive} (Xiong), and the collapsed Gibbs sampler from \cite{bennett2023fitting} (GSDPMM), all of which are established techniques for estimating sparse Markov models or related structures. As a baseline, we also compare with $k$-means clustering with several several values of $k$, and then selecting the optimal model using BIC criterion. These comparisons provide insight into the relative accuracy and robustness of our method across a variety of scenarios.

Since our method does not require the number of clusters or their labels to be pre-specified, direct computation of the misclassification rate is not feasible. Instead, we use a standard metric to evaluate clustering accuracy: the Adjusted Rand Index (ARI) defined by \cite{hubert1985comparing}, which measures the similarity between the estimated and true clusterings. A formal definition of ARI is provided in Appendix~\ref{app:ari}.

We focus especially on how different weight choices influence ARI. Previous studies, such as \cite{chi2015splitting} and \cite{sun2021convex}, have shown that using sparse weights improves both clustering accuracy and computational efficiency. We consider both dense and sparse weight matrices and assess their impact on clustering quality. We also use the pathwise approach of \cite{weylandt2020dynamic} to dynamically trace the solution path and select the optimal $\lambda$ based on BIC. Each simulation scenario is replicated 1,000 times to compute the mean ARI and standard error. We also report the empirical probability of perfect recovery—i.e., the proportion of replicates where ARI equals 1.

      \subsection{Simulation Set-up 1}
We take $|\Sigma| = 4$, mimicking DNA sequence analysis. The order of the chain is $m = 2$. The $16$ possible histories are equally divided into $4$ groups of $4$ elements each. For each group $C_i$, we generate independent $Z_{C_i, \ell} \sim \text{Unif}(0,1)$, and then define group-wise transition probabilities via a Dirichlet distribution with parameter vector $(e^{Z_{C_i,1}}, \ldots, e^{Z_{C_i,4}})$.

For convex clustering, we begin with uniform weights $w_{i,j} = 1$ for all pairs. Next, we consider sparse weights based on estimated transition probability distances. Following \cite{chi2015splitting}, we use a $k$-nearest neighbor (k-NN) approach with Gaussian kernel: $w_{i,j} = \exp(-\phi \| \hat{\boldsymbol\pi}_i - \hat{\boldsymbol\pi}_j \|_2^2) \cdot l^k_{i,j}$, where $l^k_{i,j}$ indicates if $i$ is among the $k$-nearest neighbors of $j$ or vice versa. We use $\phi = 100$ and test $k = 5$ and $k = 3$. We also explore two alternative weights: one using $l_\infty$ distance in the same form;
another using Weylandt's pathwise method to generate the full solution path and select the optimal model via BIC. We summarize the results in the following figure (\ref{fig:sim1_ari_cvx}). We also present the results in details in the table (\ref{tab:sim_1_cvx}) in the appendix.

\begin{figure}[H]
    \centering
    \begin{subfigure}[t]{0.49\textwidth}
        \begin{overpic}[width=\linewidth]{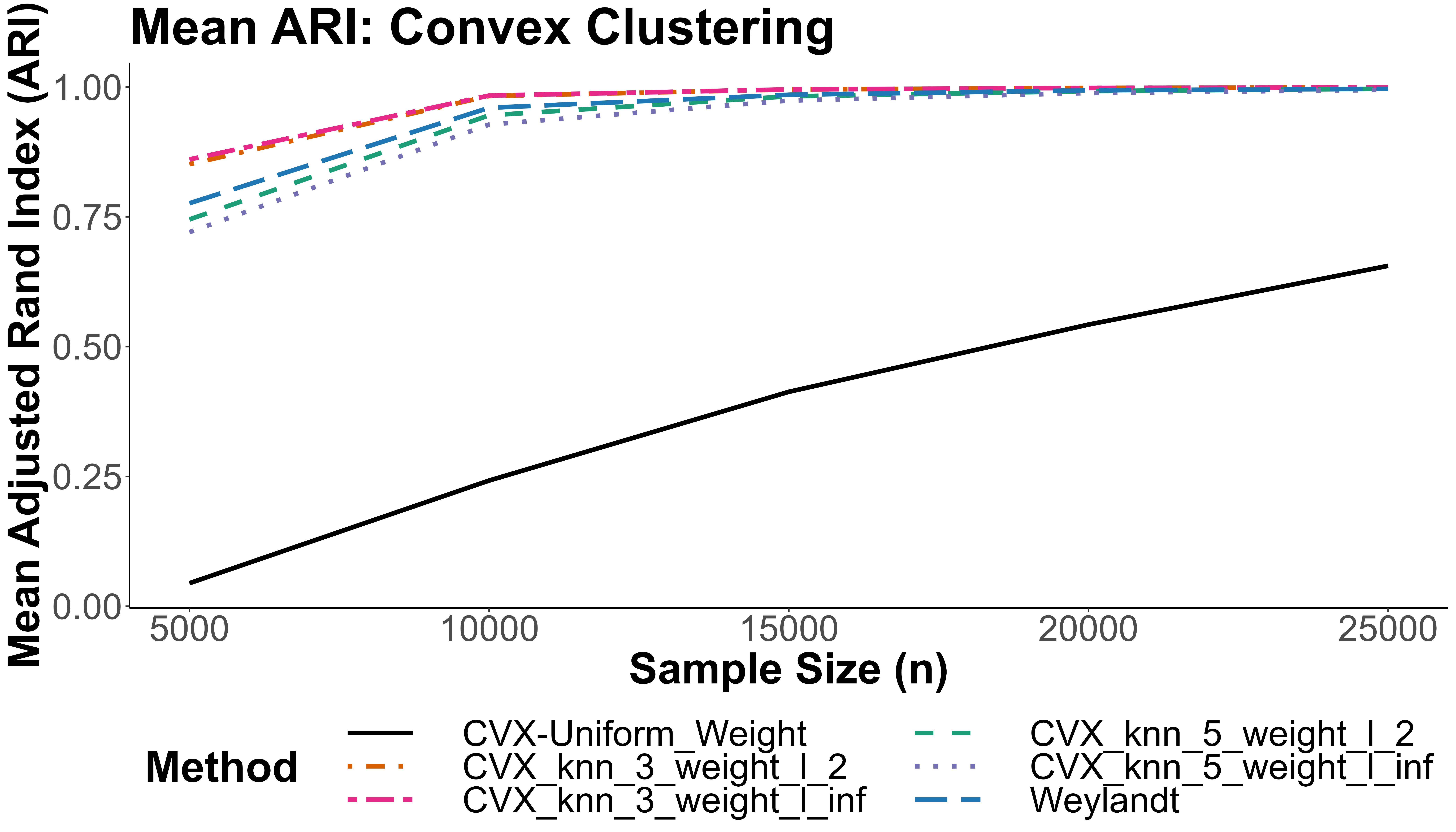}
            \put(-8,55){\textbf{(a)}}
        \end{overpic}
    \end{subfigure}
    \hfill
    \begin{subfigure}[t]{0.49\textwidth}
        \begin{overpic}[width=\linewidth]{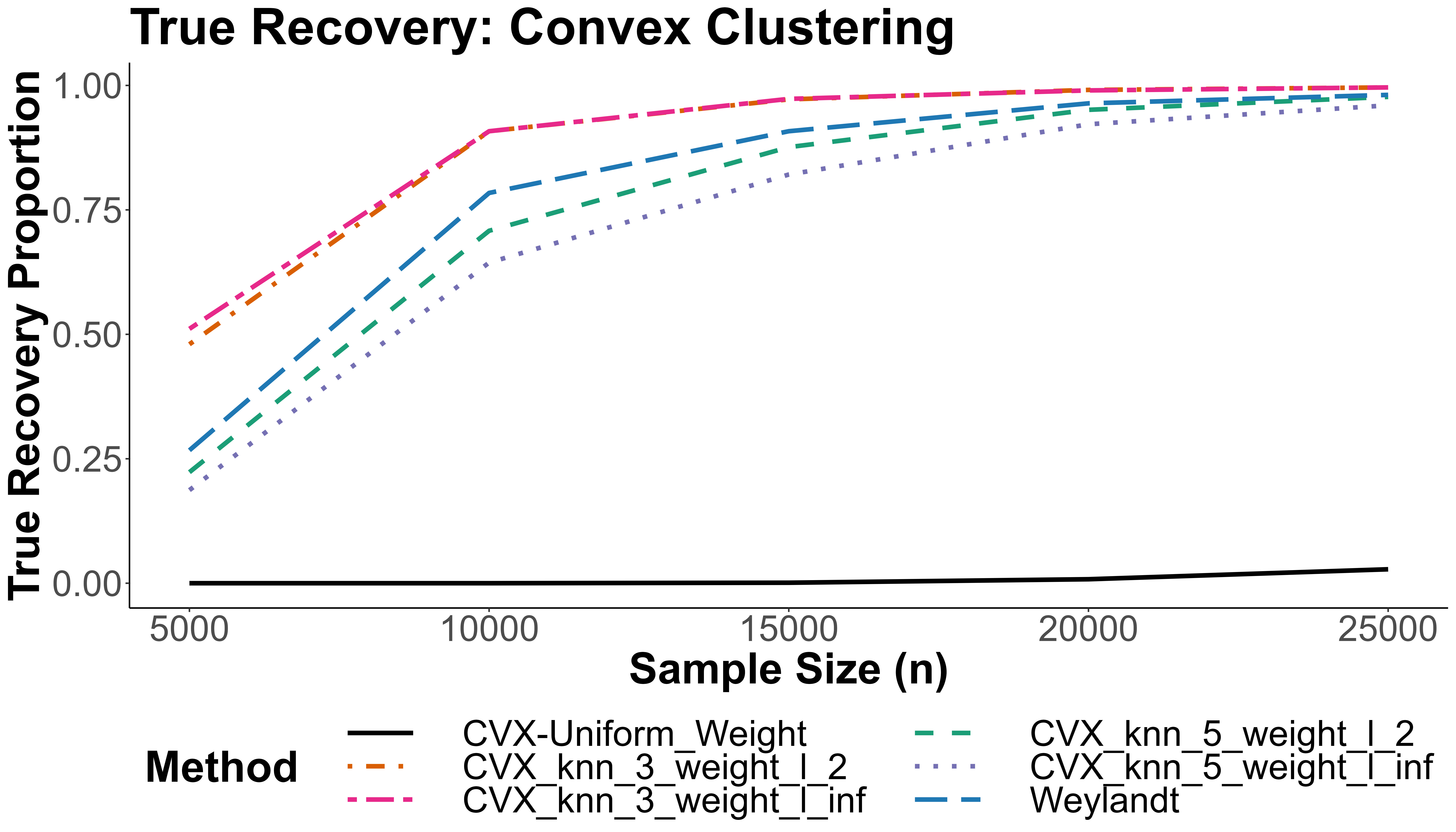}
            \put(-8,55){\textbf{(b)}}
        \end{overpic}
    \end{subfigure}
    \caption{Comparison of performances of convex clustering for different weight choices and Weylandt's pathwise method. The prefix ``CVX" indicates convex clustering, with possible weight choices as the suffix.}
    \label{fig:sim1_ari_cvx}
\end{figure}
\FloatBarrier

From the figure (\ref{fig:sim1_ari_cvx}), we observe that the uniform weights perform poorly in terms of ARI and perfect recovery, especially for smaller $n$. In contrast, sparse weights with $k=3$ significantly outperform $k=5$ and the pathwise method at lower sample sizes. This aligns with Corollary~\ref{cor_1}, which supports $k=3$ as optimal in balanced settings. As expected, performance improves for all weight choices as $n$ increases.

\begin{figure}[H]
    \centering
    \begin{subfigure}[t]{0.49\textwidth}
        \begin{overpic}[width=\linewidth]{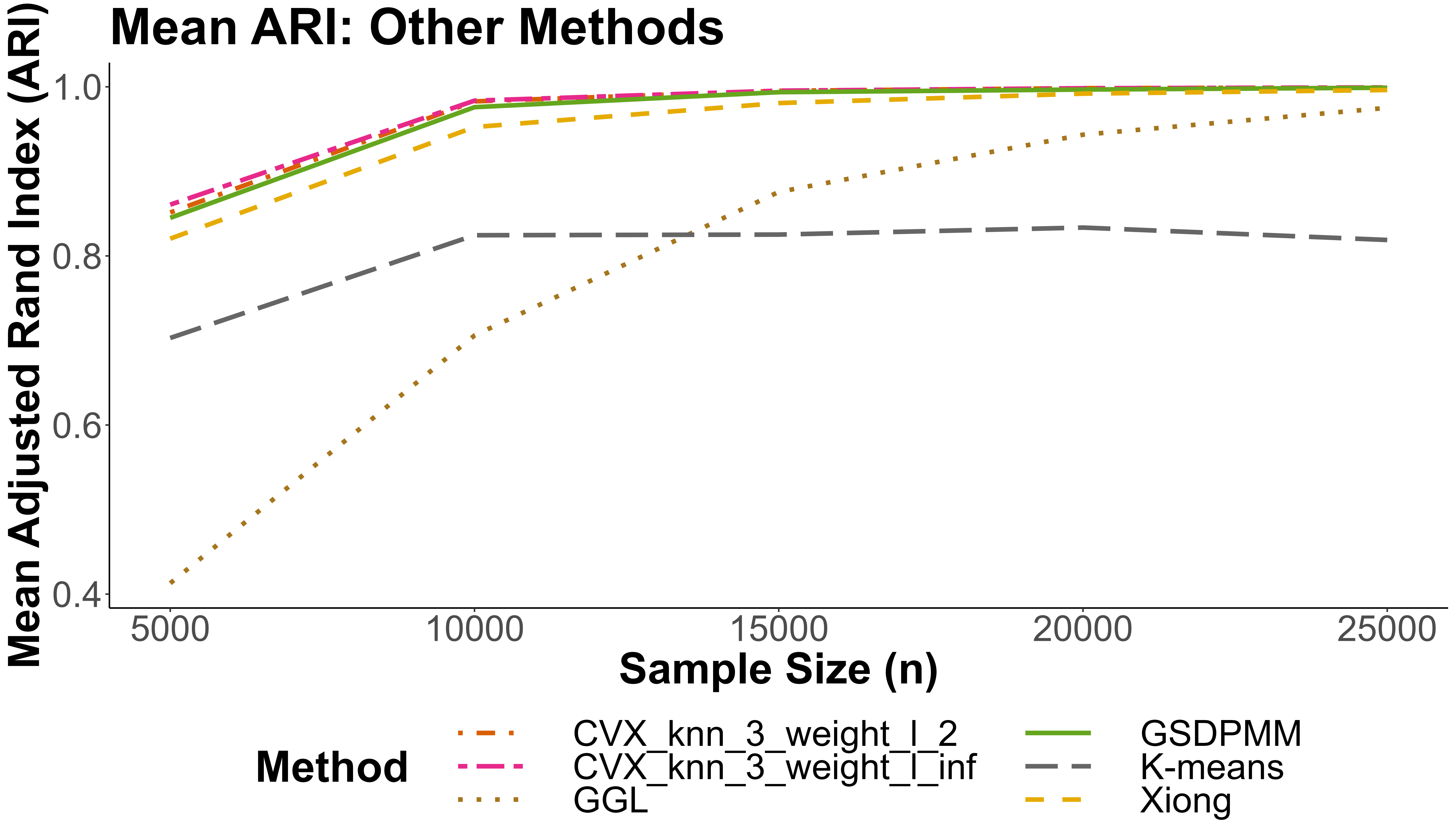}
            \put(-8,55){\textbf{(a)}}
        \end{overpic}
    \end{subfigure}
    \hfill
    \begin{subfigure}[t]{0.49\textwidth}
        \begin{overpic}[width=\linewidth]{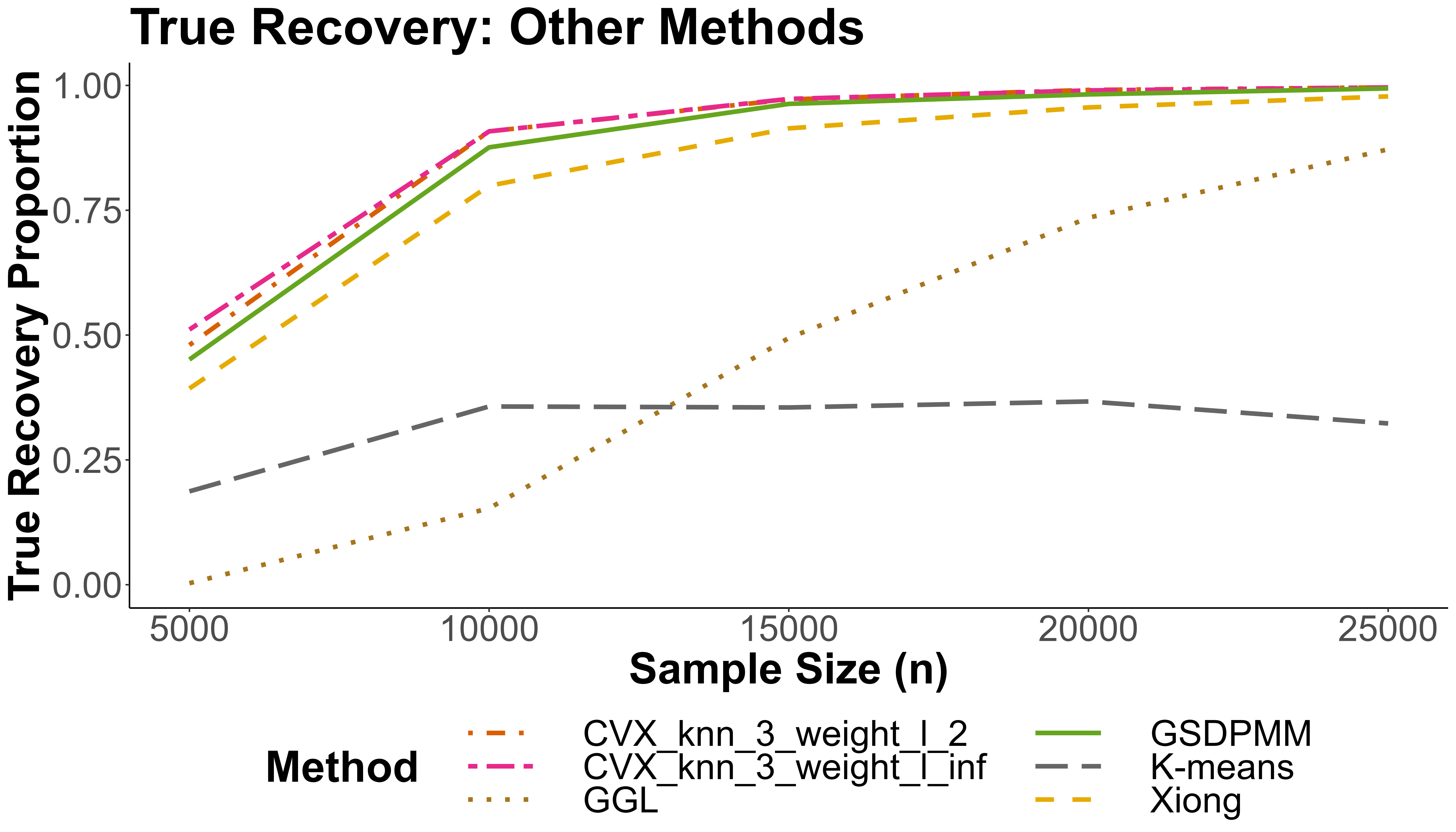}
            \put(-8,55){\textbf{(b)}}
        \end{overpic}
    \end{subfigure}
    \caption{Comparison of performances of convex clustering with competing methods of SMM fitting. The prefix ``CVX" indicates convex clustering, with possible weight choices as the suffix.}
    \label{fig:sim1_ari_other_smm}
\end{figure}
\FloatBarrier
Next, we compare our convex clustering method to the other competing methods and demonstrate the results in figure (\ref{fig:sim1_ari_other_smm}) and in the table (\ref{tab:sim_1_comparison}). Convex clustering achieves higher ARI and recovery probability than all competitors, especially for small $n$. Among competitors, GSDPMM is the closest in performance, followed by Xiong and GGL, both of which underperform. $K$-means clustering performs the worst across all sample sizes, demonstrating the advantages of convex clustering for SMM estimation.

This setup achieves near-perfect recovery when $n \geq 10,000$. However, theoretical results suggest that well-separated centroids enable high recovery even for small $n$. Here, with 4 groups ($m=2$), the minimum centroid separation is $0.129$ in $l_2$ and $0.108$ in $l_\infty$. The next simulation tests more well-separated settings.

		\subsection{Simulation Set-up 2}
		Here we take $|\Sigma|=4$ and $m=3$. We divide this $64$ triplets into four groups of sizes $18,18,15$ and $13$. For the $\alpha^{th}$ group, $R_{\alpha,\alpha}=0.7, R_{\alpha,\beta}=0.1$, $\alpha=1,2,3,4$, $\beta=1,2,3,4$, $\alpha\neq \beta$. We test three weight strategies:
        \begin{itemize}
            \item[1.] $l_2$ distance with Gaussian kernel and $k=15$,
            \item[2.] $l_\infty$ distance with exponential kernel and $\phi = 10$, i.e. $w_{i,j}=\exp^{-\phi\lVert \hat{\boldsymbol\pi}_i-\hat{\boldsymbol\pi}_j\rVert_{\infty}}l^{k(\infty)}_{i,j}$, 
            \item[3.] $l_1$ distance with exponential kernel.
        \end{itemize}
We also include Weylandt’s method, as well as all competitor methods used in Setup 1. Results for $n = 1000$ and $n = 2000$ are shown in the table (\ref{tab:ari_true_multirow_vlines}).   
\begin{table}[ht]
\centering

\begin{tabular}{r|l|cccccccc}
\toprule
$n$ & \textbf{Method} & CVX-$\ell_2$ & CVX-$\ell_\infty$ & CVX-$\ell_1$ & Weylandt & GSDPMM & Xiong & GGL & K-means \\
\midrule
\multirow{2}{*}{1000}
  & ARI        & 0.816  & 0.908 & 0.893 & 0.904 & \textbf{0.916} & 0.637 & 0.265 & 0.883 \\
  & True Rec.    & 0.000 & 0.104 & 0.030 & \textbf{0.108} & 0.104 & 0.000 & 0.000 & 0.104 \\
\midrule
\multirow{2}{*}{2000}
  & ARI        & 0.954 & 0.983 & 0.979 & 0.983 & \textbf{0.984} & 0.786 & 0.634 & 0.948 \\
  & True Rec.    & 0.140 & 0.638 & 0.468 & 0.660 & \textbf{0.683} & 0.013 & 0.000 & 0.538 \\
\bottomrule
\end{tabular}

\caption{Comparison of Adjusted Rand Index (ARI) and Probability of True Recovery across methods and sample sizes. The best performer for each $n$ and each metric (ARI or True Recovery) is shown in bold.}
\label{tab:ari_true_multirow_vlines}
\end{table}

		From the experiment, we can infer that the convex clustering with weight involving $l_2$ distance in the Gaussian kernel performs poorly compared to $l_\infty$ or $l_1$ distance. Using $l_\infty$ distance is especially effective in such scenarios, as it measures the maximum element-wise distance between two estimated transition probability vectors. We are then able to separate out two vectors that are not likely to be in the same cluster. On the other hand, the competing methods like GSDPMM, k-means or Weylandt's approach are also as good as convex clustering with $l_\infty$ distance kernel, sometimes providing a slightly better result for $n=2000$. Methods by \cite{xiong2016recursive} or \cite{garcia2011minimal} still perform worse than these methods for both sample size. This demonstrates that for well-separated clusters, we can actually use Weylandt's approach to determine the optimal $\lambda$ to get as good result as GSDPMM in much shorter time.

		\section{Real Data Analysis}\label{sec_data analysis}
		In recent years, several viral outbreaks have posed serious public health challenges. Among them, the most widespread has been the COVID-19 pandemic, caused by SARS-CoV-2, which originated in Wuhan, China in late 2019 and spread rapidly across the globe. As of April 2022, there have been over 510 million confirmed cases and more than 6.2 million deaths worldwide. Symptoms commonly include fever, headache, fatigue, and respiratory distress, as reported by \cite{wu2020characteristics}. Most cases are mild, with only 8--10\% requiring hospitalization and a fatality rate of around 1.5\%---lower than the 10\% fatality observed during the 2002 SARS-CoV-1 outbreak and much lower than the 35\% observed in the 2012 MERS-CoV outbreak in the Arabian Peninsula. Despite its lower mortality, SARS-CoV-2’s mild symptoms and high transmissibility have made it far more difficult to contain.

Diagnostic tools such as RT-PCR and rapid antigen tests have played a central role in pandemic response by identifying viral genetic material from saliva or mucus samples. While complete viral genomes can be matched to reference sequences for strain identification, in practice, clinical samples are often noisy, degraded, or partially observed---due to co-infections, sequencing limitations, or contamination. As a result, the problem of identifying the underlying virus from a \textit{partial genome} becomes statistically challenging.

This difficulty is exacerbated when multiple viruses with similar symptoms co-circulate in the population. In the Indian subcontinent, for instance, Dengue virus is endemic during summer and fall, and shares many symptoms with COVID-19. Hepatitis B also presents with fever and fatigue, making clinical differentiation harder. In critical cases, incorrect identification may lead to inappropriate treatment or fatal outcomes.

In this work, we propose a sparse Markov model (SMM)-based classification framework to identify viruses from partially observed DNA sequences. The approach is designed to be statistically efficient, interpretable, and scalable, as demonstrated in our simulation studies. We now apply this methodology to a real dataset of 500 viral genomes spanning SARS-CoV-2, MERS-CoV, Dengue, and Hepatitis B, to assess performance under various levels of sequence truncation.

		\subsection{Data Description}
		We collected genome sequences for 500 individuals from the \href{https://www.ncbi.nlm.nih.gov/labs/virus/vssi/#/virus?SeqType_s=Nucleotide}{NCBI} database, including 200 SARS-CoV-2, 50 MERS, 100 Dengue, and 150 Hepatitis B cases. The NCBI database also provides reference genome sequences, which represent canonical structures for each virus. These were used for model training. The lengths of the reference genomes are 29,903 (SARS-CoV-2), 30,119 (MERS), 10,735 (Dengue), and 3,542 (Hepatitis B).

MERS samples were collected from the 2012 outbreak in the Arabian Peninsula. Dengue and Hepatitis B samples span multiple countries and time periods over the last three decades. For SARS-CoV-2, we intentionally sampled 50 sequences from each of four critical time points: April 2020, September 2020, January 2021, and April 2021. These correspond to distinct waves or major strain transitions—such as the initial global spread, India’s first wave, the Beta wave in the U.S., and the emergence of the Delta variant in Asia. This strategy ensures temporal diversity and better representation of major SARS-CoV-2 variants.

		\subsection{Method}
        
	 As mentioned earlier, our goal is to classify viruses based on partially retrieved genome sequences. To achieve this, we first model the four reference genome sequences using our proposed SMM approach with convex clustering. Next, we randomly select a continuous segment of the genome sequence for each sample, and then compute the likelihood of that segment under each of the $4$ reference models. Suppose the $i^{th}$ model is denoted by $\hat{P}_i$, $i=1,2,3,4$. For any given sequence $x=x_1x_2\ldots x_n$, likelihood of $x$ for each model is 
		$$
		L_i(x)=\hat{P}_i\Big(\Tilde{X}_m^{(m)}=\Tilde{x}_m^{(m)}\Big)\prod_{t=m+1}^{n}\hat{P}_i\Big(X_{t+1}=x_{t+1}\Big|\Tilde{X}_t^{(m)}=\Tilde{x}_t^{(m)}\Big).
		$$
		We then classify $x$ to $\arg\max_{i=1,\ldots,4}L_i(x)$. Note that the transition probabilities inside the product term are estimated from the fitted model, along with $\hat{P}_i\Big(\Tilde{X}_m^{(m)}=\Tilde{x}_m^{(m)}\Big)$. Thus, we expect that the true virus can be classified from a moderately large segment of the full RNA sequence of the respective viruses.

        Apart from our proposed method using convex clustering, we also apply other SMM fitting methods in this setup. We apply GSDPMM, Xiong, GGL and K-means to estimate the partitions of the $m$-tuples, and their corresponding transition probability vectors. Similarly, we compute the likelihood of a DNA sequence under each reference model, and classify the DNA sequence to the most likely virus. 
        
        Depending on the assumption on the order of the Markov chains, we fit two different models in our analysis as follows.
        
    \paragraph{Model 1.} In the first model, we fit SMM-s of order $m=4$ for SARS and MERS, while for the other two viruses, we use $m=3$. For convex clustering, the form of the weights are $w_{i,j}=\exp^{-\phi\lVert \hat{\boldsymbol\pi}_i-\hat{\boldsymbol\pi}_j\rVert_{\infty}^2}l^k_{i,j}$ where $l^k_{i,j}$ is the indicator function that $\hat{\boldsymbol\pi}_i$ belongs to $k$ nearest neighbour of $\hat{\boldsymbol\pi}_j$ or vice versa in terms of $l_{\infty}$ distance, for some $\phi=100$. For $m=4$, we set the number of nearest neighbors to $k=20$, and for $m=3$, we set $k=5$. We fit the four competing methods with their default parameter settings.
    \paragraph{Model 2.} In this model, we fit SMM with $m=3$ for all four viruses. We take the same weight choice with $\phi=100$ and $k=5$ for convex clustering, and default parameters for the competing methods. 

    From the samples, we randomly choose segments of length $100\epsilon\%$, and compute the likelihoods under $4$ models to classify it to the most likely class of virus. We consider three values of $\epsilon$: $0.05$, $0.1$, and $0.25$. We compute the overall mis-classification rates for all three cases, i.e. the proportion of virus that are wrongly classified in each case.
		\subsection{Results}

		\subsubsection{Model 1}
\begin{figure}[H]
    \centering
    \includegraphics[width=\textwidth]{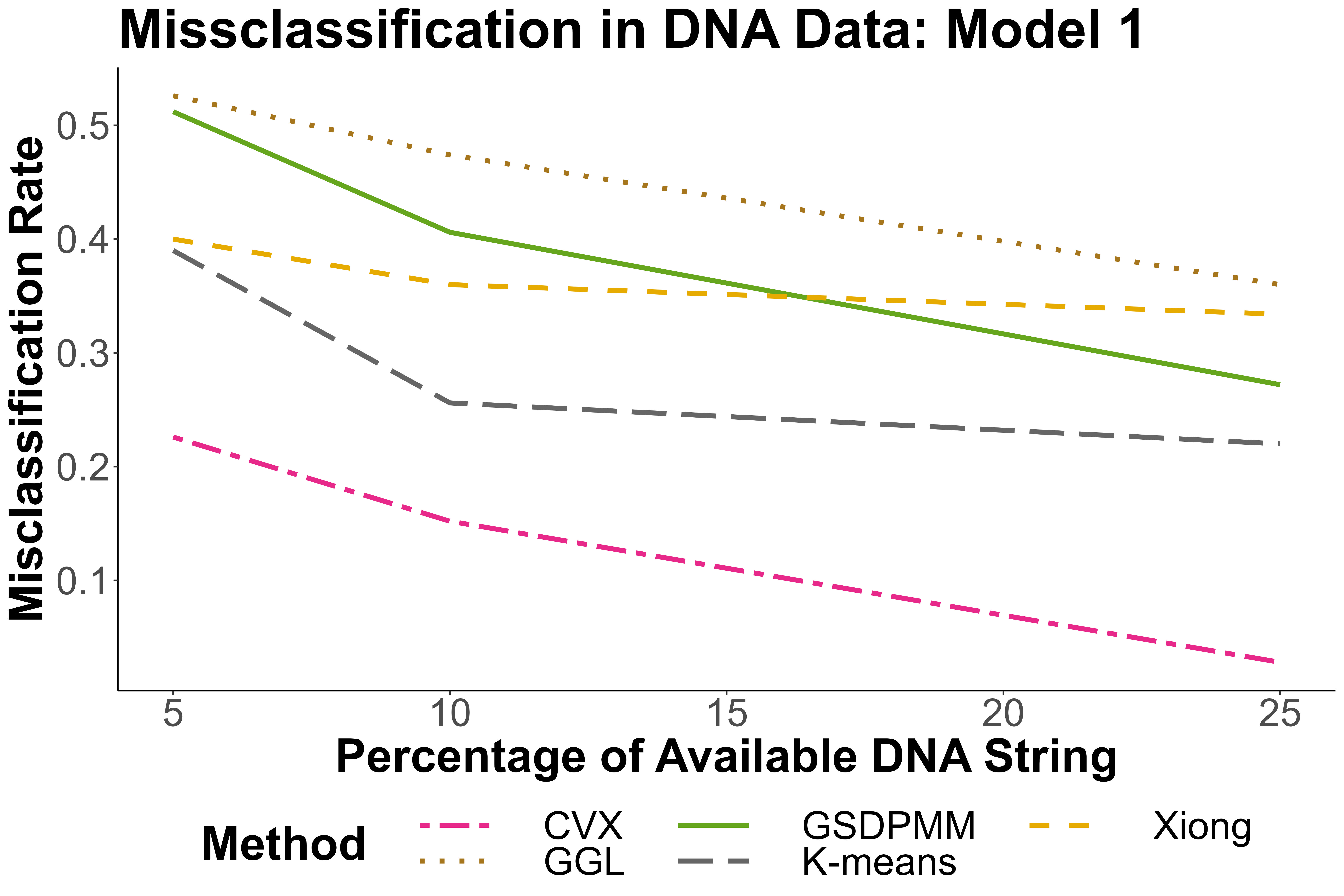}
    \caption{Comparison of performances of convex clustering (CVX) to different SMM-fitting method under Model 1 for the virus classification problem.}
    \label{fig:real_data_model1}
\end{figure}

		Table (\ref{table:cluster_size}) shows the number and sizes of clusters for each reference model obtained via convex clustering. We also present the class-wise counts of the samples from a particular species in the $4\times 4$ confusion matrices in the following table (\ref{table:model1}) and report the mis-classification rates. Figure (\ref{fig:real_data_model1}) compares the mis-classification rates of convex clustering with the competing methods.
		
		\begin{table}[htbp!]
			\centering
			\begin{tabular}{|c|c|c|}
				\hline
				{\color[HTML]{333333} \textbf{Virus}} & {\color[HTML]{333333} \textbf{\begin{tabular}[c]{@{}c@{}}Number of\\ Clusters\end{tabular}}} & {\color[HTML]{333333} \textbf{Cluster Size}}                                                                                       \\ \hline
				\hline
				{\color[HTML]{333333} Covid 19}       & {\color[HTML]{333333} 28}                                                                    & {\color[HTML]{333333} \begin{tabular}[c]{@{}c@{}}155, 50, 10, 7, 5, 4, \\ 2 (3 times), 1 (19 times)\end{tabular}}                        \\ \hline
				{\color[HTML]{333333} MERS}           & {\color[HTML]{333333} 30}                                                                    & {\color[HTML]{333333} \begin{tabular}[c]{@{}c@{}}141, 60, 6, 5 (3 times),\\ 3 (3 times), 2 (4 times),\\ 1 (17 times)\end{tabular}} \\ \hline
				{\color[HTML]{333333} Dengue}         & {\color[HTML]{333333} 7}                                                                     & {\color[HTML]{333333} 24, 16, 14, 4, 3, 2, 1}                                                                                      \\ \hline
				{\color[HTML]{333333} Hepatitis B}    & {\color[HTML]{333333} 14}                                                                    & {\color[HTML]{333333} 41, 8, 4, 1 (11 times)}                                                                                      \\ \hline
			\end{tabular}
			\caption{Number of clusters and size of each cluster obtained in Model 1.}
			
			\label{table:cluster_size}
		\end{table}

		\begin{table}[htbp!]
			\centering
			\begin{tabular}{|c|cccc|c|}
				\hline
				\backslashbox{\textbf{Observed}}{\textbf{Fitted}}      & \textbf{\begin{tabular}[c]{@{}c@{}}SARS-\\ Cov-2\end{tabular}} & \textbf{MERS} & \textbf{Dengue} & \textbf{\begin{tabular}[c]{@{}c@{}}Hepatitis \\ B\end{tabular}} & \textbf{Total} \\ \hline
				\textbf{SARS-Cov-2 } & {\color[HTML]{32CB00}\textbf{185}}                                                            & 0            & 4               & 11                                                               & 200            \\
				\textbf{MERS }       & 0                                                              & {\color[HTML]{036400} \textbf{50}}             & 0               & 0                                                               & 50             \\
				\textbf{Dengue }     & 0                                                              & 0             & {\color[HTML]{036400} \textbf{100}}             & 0                                                               &         100    \\
				\textbf{Hepatitis B} & 17                                                             & 44            & 37              & {\color[HTML]{CB0000} \textbf{52}}                                                              & 150            \\ \hline
			\end{tabular}
			
			\begin{tabular}{|c|cccc|c|}
				\hline
				\backslashbox{\textbf{Observed}}{\textbf{Fitted}}      & \textbf{\begin{tabular}[c]{@{}c@{}}SARS-\\ Cov-2\end{tabular}} & \textbf{MERS} & \textbf{Dengue} & \textbf{\begin{tabular}[c]{@{}c@{}}Hepatitis \\ B\end{tabular}} & \textbf{Total} \\ \hline
				\textbf{SARS-Cov-2 } & {\color[HTML]{32CB00}\textbf{193}}                                                            & 0          & 0               & 7                                                               & 200            \\
				\textbf{MERS }       & 0                                                              & {\color[HTML]{036400} \textbf{50}}            & 0               & 0                                                               & 50             \\
				\textbf{Dengue }     & 0                                                              & 0             & {\color[HTML]{036400} \textbf{100}}             & 0                                                               & 100            \\
				\textbf{Hepatitis B} & 6                                                             & 39            & 24              & {\color[HTML]{FE7600} \textbf{81}}                                                              & 150            \\ \hline
			\end{tabular}
			
			\begin{tabular}{|c|cccc|c|}
				\hline
				\backslashbox{\textbf{Observed}}{\textbf{Fitted}}      & \textbf{\begin{tabular}[c]{@{}c@{}}SARS-\\ Cov-2\end{tabular}} & \textbf{MERS} & \textbf{Dengue} & \textbf{\begin{tabular}[c]{@{}c@{}}Hepatitis \\ B\end{tabular}} & \textbf{Total} \\ \hline
				\textbf{SARS-Cov-2 } & {\color[HTML]{32CB00}\textbf{194}}                                                             & 0            & 0               & 6                                                               & 200            \\
				\textbf{MERS }       & 0                                                              & {\color[HTML]{036400} \textbf{50}}             & 0               & 0                                                               & 50             \\
				\textbf{Dengue }     & 0                                                              & 0             & {\color[HTML]{036400} \textbf{100}}              & 0                                                               & 100            \\
				\textbf{Hepatitis B} & 0                                                             & 8            & 0              & {\color[HTML]{32CB00}\textbf{142}}                                                              & 150            \\ \hline
			\end{tabular}
			\caption{Confusion Matrices for $\epsilon=0.05$, $0.1$ and $0.25$ respectively with mis-classification rates $22.6\%$, $15.2\%$ and $2.8\%$ in \textbf{Model 1} with \textbf{Convex Clustering}.}
			\label{table:model1}
		\end{table}

\FloatBarrier
		
		\subsubsection{Model 2}
\begin{figure}[H]
    \centering
    \includegraphics[width=\textwidth]{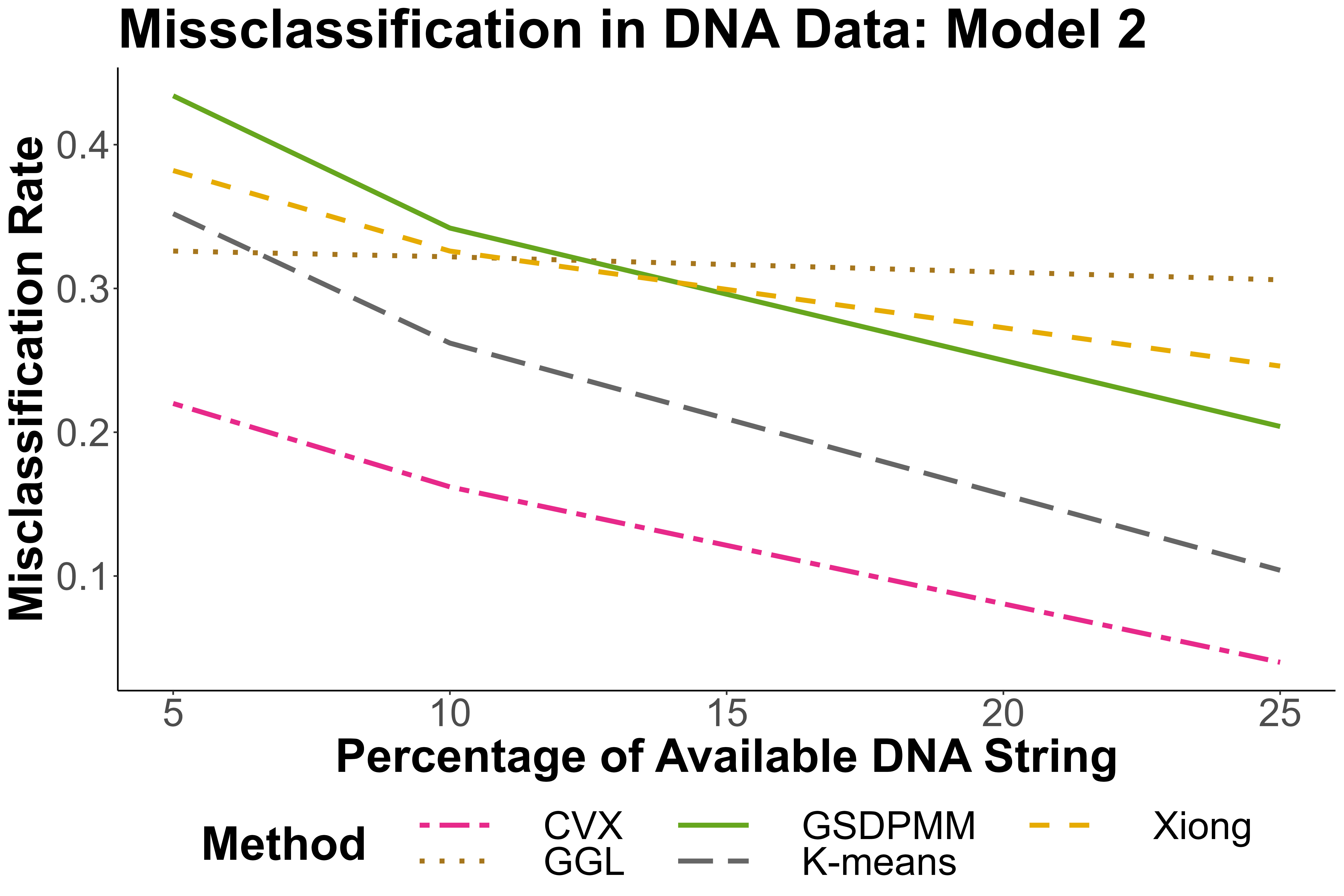}
    \caption{Comparison of performances of convex clustering (CVX) to different SMM-fitting method under Model 2 for the virus classification problem.}
    \label{fig:real_data_model2}
\end{figure}

		Now we present the clustering performance for the second model using convex clustering, where all fitted SMM have order $m=3$ in the table (\ref{table:model2}). In figure (\ref{fig:real_data_model2}), we have compared the performance of convex clustering with other methods.
		\begin{table}[htbp!]
			\centering
			
			\begin{tabular}{|c|cccc|c|}
				
				\hline
				\backslashbox{\textbf{Observed}}{\textbf{Fitted}}      & \textbf{\begin{tabular}[c]{@{}c@{}}SARS-\\ Cov-2\end{tabular}} & \textbf{MERS} & \textbf{Dengue} & \textbf{\begin{tabular}[c]{@{}c@{}}Hepatitis \\ B\end{tabular}} & \textbf{Total} \\ \hline
				\textbf{SARS-Cov-2 } & {\color[HTML]{32CB00}\textbf{177}}                                                            & 8            & 4               & 11                                                               & 200            \\
				\textbf{MERS }       & 1                                                              & {\color[HTML]{036400} \textbf{49}}             & 0               & 0                                                               & 50             \\
				\textbf{Dengue }     & 0                                                              & 0             & {\color[HTML]{036400} \textbf{100}}             & 0                                                               &         100    \\
				\textbf{Hepatitis B} & 11                                                             & 30            & 45              & {\color[HTML]{CB0000} \textbf{64}}                                                              & 150            \\ \hline
				
			\end{tabular}
			
			
			\begin{tabular}{|c|cccc|c|}
				\hline
				\backslashbox{\textbf{Observed}}{\textbf{Fitted}}      & \textbf{\begin{tabular}[c]{@{}c@{}}SARS-\\ Cov-2\end{tabular}} & \textbf{MERS} & \textbf{Dengue} & \textbf{\begin{tabular}[c]{@{}c@{}}Hepatitis \\ B\end{tabular}} & \textbf{Total} \\ \hline
				\textbf{SARS-Cov-2 } & {\color[HTML]{32CB00}\textbf{187}}                                                            & 6          & 0               & 7                                                               & 200            \\
				\textbf{MERS }       & 0                                                              & {\color[HTML]{036400} \textbf{50}}            & 0               & 0                                                               & 50             \\
				\textbf{Dengue }     & 0                                                              & 0             & {\color[HTML]{036400} \textbf{100}}             & 0                                                               & 100            \\
				\textbf{Hepatitis B} & 4                                                             & 37            & 27              & {\color[HTML]{FE7600} \textbf{82}}                                                              & 150            \\ \hline
			\end{tabular}
			
			\begin{tabular}{|c|cccc|c|}
				\hline
				\backslashbox{\textbf{Observed}}{\textbf{Fitted}}      & \textbf{\begin{tabular}[c]{@{}c@{}}SARS-\\ Cov-2\end{tabular}} & \textbf{MERS} & \textbf{Dengue} & \textbf{\begin{tabular}[c]{@{}c@{}}Hepatitis \\ B\end{tabular}} & \textbf{Total} \\ \hline
				\textbf{SARS-Cov-2 } & {\color[HTML]{32CB00}\textbf{190}}                                                             & 2            & 0               & 8                                                               & 200            \\
				\textbf{MERS }       & 0                                                              & {\color[HTML]{036400} \textbf{50}}             & 0               & 0                                                               & 50             \\
				\textbf{Dengue }     & 0                                                              & 0             & {\color[HTML]{036400} \textbf{100}}              & 0                                                               & 100            \\
				\textbf{Hepatitis B} & 0                                                             & 10            & 0              & {\color[HTML]{32CB00}\textbf{140}}                                                              & 150            \\ \hline
			\end{tabular}
			\caption{Confusion Matrices for $\epsilon=0.05$, $0.1$ and $0.25$ respectively with mis-classification rates $22\%$, $16\%$ and $4\%$ in \textbf{Model 2} with \textbf{Convex Clustering}.}
			\label{table:model2}
		\end{table}

\FloatBarrier
		\subsection{Discussion}

In statistical analysis, larger sample sizes typically yield more reliable inference. In our experiment, when only a small portion of the genome is retained, classification performance suffers—particularly for Hepatitis B. At $\epsilon = 0.05$, the average retained segment is just 170 bases for Hepatitis B, compared to approximately 500 for Dengue and 1500 for MERS and SARS-CoV-2. This disparity explains the elevated error rate for Hepatitis B in low-$\epsilon$ settings. As the proportion of the sequence increases, accuracy improves substantially: misclassification rates fall from 22.8\% to 3.2\% as $\epsilon$ increases from $0.05$ to $0.25$. Even for Hepatitis B, error drops sharply when moderately long segments are retained. For the other three viruses, misclassifications are rare across all $\epsilon$ levels.

Figures (\ref{fig:real_data_model1}) and (\ref{fig:real_data_model2}) confirm that convex clustering consistently outperforms other methods (GSDPMM, Xiong, GGL) across all settings. While all methods benefit from longer segments, convex clustering maintains a significant edge. At $\epsilon = 0.25$, competing methods still exceed 30\% error, while convex clustering achieves rates of only 4\% and 2.8\% in Models 1 and 2, respectively.

Choosing the appropriate Markov order $m$ is a central issue in fitting SMMs. Higher $m$ values capture richer dependencies but risk data sparsity. For instance, setting $m=10$ would result in many histories with zero observations. To balance this trade-off, we select $m$ based on sequence length—using $m=4$ for the long genomes of SARS-CoV-2 and MERS (near 30,000 bases) and $m=3$ for Dengue and Hepatitis B. This empirically grounded choice ensures that most $m$-tuples are well-represented in the data.

Beyond numerical considerations, there is a biological rationale for choosing $m \geq 3$. In DNA and RNA, three consecutive bases form a codon, which encodes either an amino acid or a stop signal during protein synthesis. Although there are 64 possible codons, they map to only 20 amino acids, with some redundancy, and a few codons act as stop signals. This triplet-based structure supports using SMMs of order at least 3 to reflect functional genomic units.

Finally, even though sequence snippets were sampled randomly from the full genomes, accurate classification was often possible. This suggests that distinctive sequence features are dispersed throughout the genome. Short segments may capture shared motifs or conserved regions (e.g., spike proteins), which can lead to mis-classification. But as sequence length increases, virus-specific patterns dominate. For example, MERS and Dengue were never misclassified by convex clustering, even at $\epsilon = 0.05$. This highlights our model’s ability to extract robust features from local sequence statistics, while the existing SMM-fitting methods fail to capture these features.

		\section{Summary} \label{sec_conclusion}

We proposed a convex clustering-based method for fitting sparse Markov models (SMMs), which adaptively groups transition distributions using a data-driven fusion penalty. Our approach is fully unsupervised, requiring no prior knowledge of the number of clusters or transition probabilities, and is backed by strong theoretical guarantees. In particular, we show that the true clustering structure can be consistently recovered as sample size increases, enabling efficient dimension reduction while maintaining model interpretability.

Through extensive simulation studies, we benchmarked the method against several competing approaches—GSDPMM, Xiong et al., GGL, and K-means—and found that convex clustering consistently delivers higher clustering accuracy, especially in noisy or high-dimensional settings. The method remains stable across a range of signal strengths, Markov orders, and weighting schemes, making it well-suited for general-purpose sequence modeling. The real data application to viral genomes illustrates the method’s practical relevance. Even with severely truncated sequences (e.g., retaining only 5–10\% of the genome), the method accurately classifies viruses such as SARS-CoV-2, MERS, and Dengue, outperforming other methods by a substantial margin.

In summary, our method offers a statistically rigorous and computationally efficient framework for clustering and classifying sequence data under sparsity constraints. Its robustness to noise, model flexibility, and biological interpretability position it as a promising tool for applications in genomics, large language modeling, and beyond.

\if1\blind
{
\section*{Acknowledgments}
The authors would like to thank Dr. Iris Bennett for providing helpful code and computational suggestions, and two anonymous referees for their valuable suggestions for the improvement of the paper. They also thank the Department of Statistics, North Carolina State University for providing computing resources.
\section*{Code Availability}
The R code and results for the simulation studies and real data analysis are available at: \url{https://github.com/tuhinmajumderstat/SMM-fit-Convex-Clustering}.
}
\fi 
\if0\blind
{
\section*{Acknowledgments}
The authors would like to thank the collaborators and two anonymous referees for their valuable suggestions for the improvement of the paper.
\section*{Code Availability}
The R code and results for the simulation studies and real data analysis are uploaded in a `.zip' file as a supplementary document for review. 
}
\fi 
\section*{Data Availability}
The reference and the complete genome sequences have been downloaded from the \href{https://www.ncbi.nlm.nih.gov/labs/virus/vssi/#/virus?SeqType_s=Nucleotide}{NCBI} database. The exact datasets we have used can be found in this \href{https://doi.org/10.7910/DVN/X54HYA}{Harvard Dataverse} repository. They are also available in the supplementary files with the paper.
		\appendix
			
			\section{Proof of Theorems}
			\subsection{Proof of Theorem \ref{thm_prob_dist}} \label{proof1}
			
			(a) For notational simplicity, we write $b^*_{i,a}(\lambda)$ as $b^*_{i,a}$. Let $$
			R(\mathbf{B},\mathbf{W})= \dfrac{1}{2}\sum_{j=1}^p \lVert \hat{\boldsymbol\pi}_j - \mathbf{b}_{j} \rVert_2^2
			+\la\sum_{1\leq i<j\leq p} w_{i,j} \lVert\mathbf{b}_{i}- \mathbf{b}_{j}\rVert_2. 
			$$  Suppose $b^*_{i,a}< 0$ for some of the $(i,a)$ pairs, $i=1,2,\ldots,p$ and $a=1,2,\ldots,d$. Let, $b^{**}_{i,a}=b^*_{i,a}\mathcal{I}(b^*_{i,a}>0)$. Since $\hat{\pi}_{i,a}\ge 0$, we get $\big\lvert\hat{\pi}_{i,a}-b^{**}_{i,a}\big\rvert \le \big\lvert\hat{\pi}_{i,a}-b^{*}_{i,a}\big\rvert$. Also, $\big\lvert b^{*}_{i_1,a}-b^{*}_{i_2,a}\big\rvert\ge \big\lvert b^{**}_{i_1,a}-b^{**}_{i_2,a}\big\rvert$, since the negative elements are shrunk to $0$. Hence for any $i=1,2,\ldots,p$,
			$$
			\lVert\hat{\boldsymbol\pi}_i- \mathbf{b}^*_i \rVert_2^2 \ge \lVert\hat{\boldsymbol\pi}_i- \mathbf{b}^{**}_i \rVert_2^2; \quad \big\lVert \mathbf{b}^{*}_{i_1}-\mathbf{b}^{*}_{i_2}\big\rVert_2\ge \big\lVert \mathbf{b}^{**}_{i_1}-\mathbf{b}^{**}_{i_2}\big\rVert_2.
			$$
			Since $\mathbf{b}^{*}_{i_1}\ne \mathbf{b}^{**}_{i}$ for at least one $i$, $R(\mathbf{B}^{**},\mathbf{W})<R(\mathbf{B}^*,\mathbf{W})$, contradicting that $\mathbf{B}^*$ is the optimum solution. Hence $b^*_{i,a}\ge 0$, $\forall i=1,\ldots,p;$ $a=1,,,.d$.
			\\[.1 in]
			(b) If we initialize $\boldsymbol\Gamma^{(0)}=\boldsymbol 0$, we get $\mathbf{b}_i^{(1)}=\hat{\boldsymbol\pi}_i$, which satisfies $\sum_{a=1}^{d} b^{(1)}_{i,a}=1$. Subsequently, $\boldsymbol\gamma_l^{(1)}  = \mathcal{P}_{C_l}(\boldsymbol\gamma_l^{(0)}-\nu \mathbf{g}_l^{(1)})=(\boldsymbol\gamma_l^{(0)}-\nu \mathbf{g}_l^{(1)})\min\Big\{1,\dfrac{\lambda w_l }{\lVert\boldsymbol\gamma_l^{(0)}-\nu \mathbf{g}_l^{(1)}\rVert_2}\Big\}$, and thus 
			$\boldsymbol\gamma_l^{(1)T}\bm{1}=0$. Using  a similar argument, for any iteration $t$, $\sum_{a=1}^{d} b^{(t)}_{i,a}=1$. Hence the limiting quantity will still have the property that the sum of the elements of $b_i$ is always $1$. This completes the proof that $\mathbf{b}_i^*$ is indeed a probability distribution.
			
			\subsection{Proof of Theorem \ref{thm_cluster_recovery}}\label{Proof2} 
			Note that as $n\to\infty$, $N_{\sigma_j}\to\infty$. Let $q_{j}$ be the stationary probability of the state $\si_j$. Then, $N_{\sigma_j}/(n-m)\xrightarrow[]{p} q_j$ as $n\to\infty$; and for $j\in\C_{\alpha}$, we have
			$$
			\begin{aligned}
				&\sqrt{N_{\sigma_j}}\big(\hat{\boldsymbol\pi}_j-\bm{R}_\alpha\big)\xrightarrow[]{d} \mathcal{N}(\bm{0},\Sigma_\alpha)\\
				\implies & \sqrt{(n-m)}\big(\hat{\boldsymbol\pi}_j-\bm{R}_\alpha\big)\xrightarrow[]{d} \mathcal{N}(\bm{0},q_j\Sigma_\alpha)
			\end{aligned}
			$$
			where $\Sigma_\alpha=diag(\bm{R}_\alpha)-\bm{R}_\alpha\bm{R}_\alpha^{(T)}$.  \\
			
			The proof mainly relies on calculating the probability of $\hat{\boldsymbol\pi}_j$ and $\bm{R}_\alpha$ being close to each other for all $j\in\C_\alpha.$ Suppose $\lVert\hat{\boldsymbol\pi}_j-\bm{R}_\alpha\rVert_2<\epsilon/2$ for some $\epsilon>0$, and $\forall j\in\C_\alpha$, $\alpha=1,2\ldots,k_0$. In that case, suppose $\lVert\hat{\boldsymbol\pi}_j-\bm{R}_\alpha\rVert_2<\epsilon/2$ for some $\epsilon>0$, and $\forall j\in\C_\alpha$, $\alpha=1,2\ldots,k_0$. In that case,
			$$
			\begin{aligned}
				\lambda_{\text{min}}^{(n)} & < \dfrac{\epsilon/2}{\min\limits_{1\le\alpha\le k_0}  \min\limits_{i,j\in\C_{\alpha}}\big(p_\alpha w_{i,j}-\mu_{i,j}^{(\alpha)}\big)}\\
				\lambda_{\text{max}}^{(n)} & > \min_{1\le\alpha\le k_0} \Bigg\{\dfrac{\lVert \bm{R}_\alpha-\bm{R}_\beta\rVert_2-\epsilon}{\frac{1}{p_{\alpha}}\sum_{l\ne\alpha}w^{(\alpha,l)}+\frac{1}{p_{\beta}}\sum_{l\ne\beta}w^{(\beta,l)}}\Bigg\}.
			\end{aligned}
			$$
			Thus, for $\epsilon$ sufficiently small, $\lambda_{\text{min}}^{(n)}<\lambda_{\text{max}}^{(n)}$. We will later find a bound on how small $\epsilon$ needs to be to achieve this. 
			
			We compute a lower bound on the following probability:
			$$
			P\Big(\lVert\hat{\boldsymbol\pi}_j-\bm{R}_\alpha\rVert_2<\dfrac{\epsilon}{2};\forall j\in\C_\alpha,\forall\alpha=1,\ldots,k_0\Big).
			$$
			Note that the variance-covariance matrix $\Sigma_\alpha$ of the limiting distribution is not full rank, as we have a linear constraint on the elements of $\boldsymbol{\pi}_j$. Define $Z_j=(\hat{\pi}_{j,1}-R_{j,1},\ldots,\hat{\pi}_{j,d-1}-R_{j,d-1})^T$, and let $\Sigma_{\alpha,-d}$ be the upper $(d-1)\times (d-1)$ block of $\Sigma_\alpha$. Now,
			$$
			\begin{aligned}
				\lVert\hat{\boldsymbol\pi}_j-\bm{R}_\alpha\rVert_2^2&=\sum_{l=1}^{d-1} (\hat{\pi}_{j,l}-R_{j,l})^2+\Big(\sum_{l=1}^{d-1}(\hat{\pi}_{j,l}-R_{j,l})\Big)^2\\
				&=Z_j^TZ_j+(\bm{1}^TZ_j)^2=Z_j^T(\bm{I}+\bm{1}\bm{1}^T)Z_j.
			\end{aligned}
			$$
			Define $U_j=\sqrt{\dfrac{n-m}{q_j}}\Sigma_{\alpha,-d}^{-1/2}Z_j$. By the asymptotic normality of $\hat{\boldsymbol\pi}_j$, $\sqrt{n-m} Z_j\xrightarrow[]{d} \mathcal{N}(\bm{0},q_j\Sigma_{\alpha,-d})$, hence $U_j\xrightarrow[]{d} \mathcal{N}(\bm{0},\bm{I})$. Thus,
			$$
			\begin{aligned}
				P\Big(\lVert\hat{\boldsymbol\pi}_j-\bm{R}_\alpha\rVert_2\ge\dfrac{\epsilon}{2}\Big)&=P\Big(Z_j^T(\bm{I}+\bm{1}\bm{1}^T)Z_j\ge\dfrac{\epsilon^2}{4}\Big)=P\Big(U_j^T\Sigma_{\alpha,-d}^{1/2}(\bm{I}+\bm{1}\bm{1}^T)\Sigma_{\alpha,-d}^{1/2}U_j\ge\dfrac{(n-m)\epsilon^2}{4q_j}\Big)\\
				&=P\Big(U_j^T\bm{M}U_j\ge\dfrac{(n-m)\epsilon^2}{4q_j}\Big);\quad \bm{M}=\Sigma_{\alpha,-d}^{1/2}(\bm{I}+\bm{1}\bm{1}^T)\Sigma_{\alpha,-d}^{1/2}.
			\end{aligned}
			$$
			For a symmetric matrix matrix $\bm{M}_1$, \cite{hanson1971bound} have determined a lower bound on the tail probability of any quadratic form $U^T\bm{M}_1U$ of a sub-Gaussian random variable $U$ with mean $\bm{0}$ and variance-covariance matrix $\si^2\bm{I}$ as follows:
			\begin{equation}\label{hanson-wright}
				P\Big(U^T\bm{M}_1U\ge t+\si^2tr(\bm{M}_1)\Big)\le \exp\Big[-\min\Big(\dfrac{a_1t^2}{\si^4\lVert \bm{M}_1\rVert_F},\dfrac{a_2 t}{\si^2\lVert \bm{M}_1\rVert_{sp}}\Big)\Big]
			\end{equation}
			for some constants $a_1,a_2>0$. Here $\lVert.\rVert_F$ and $\lVert.\rVert_{sp}$ are the Frobenius and spectral norms, respectively. Applying the bound in (\ref{hanson-wright}) to our problem, we obtain, as $n\to \infty$,
			$$
			\begin{aligned}
				P\Big(\lVert\hat{\boldsymbol\pi}_j-\bm{R}_\alpha\rVert_2\ge\dfrac{\epsilon}{2}\Big)&=P\Big(U_j^T\bm{M}U_j\ge\dfrac{(n-m)\epsilon^2}{4q_j}\Big)\\
				&\le \exp\Big[-\min\Big(\dfrac{a_1((n-m)\epsilon^2-4q_jtr(\bm{M}))^2}{16q_j^2\lVert \bm{M}\rVert_F},\dfrac{a_2 ((n-m)\epsilon^2-4q_jtr(\bm{M}))}{4q_j\lVert \bm{M}\rVert_{sp}}\Big)\Big].
			\end{aligned}
			$$
			 As $n$ increases, $(n-m)^2\gg (n-m)$, and eventually for larger $n$, $\min\Big(\dfrac{a_1((n-m)\epsilon^2-4q_jtr(\bm{M}))^2}{16q_j^2\lVert \bm{M}\rVert_F},$ $\dfrac{a_2 ((n-m)\epsilon^2-4q_jtr(\bm{M}))}{4q_j\lVert \bm{M}\rVert_{sp}}\Big)$ $=\dfrac{a_2 ((n-m)\epsilon^2-4q_jtr(\bm{M}))}{4q_j\lVert \bm{M}\rVert_{sp}}$. Now,
			$$
			\begin{aligned}
				tr(\bm{M})&=tr(\Sigma_{\alpha,-d}^{1/2}(\bm{I}+\bm{1}\bm{1}^T)\Sigma_{\alpha,-d}^{1/2})=tr(\Sigma_{\alpha,-d})+tr(\bm{1}^T\Sigma_{\alpha,-d}\bm{1})\\
				&=\sum_{l=1}^{d-1} R_{\alpha,l}(1-R_{\alpha,l}) + \sum_{l=1}^{d-1} R_{\alpha,l}- \sum_{l_1=1}^{d-1} \sum_{l_2=1}^{d-1}R_{\alpha,l_1} R_{\alpha,l_2} \\
				&= \sum_{l=1}^{d-1} R_{\alpha,l}(1-R_{\alpha,l}) + \Big(\sum_{l=1}^{d-1} R_{\alpha,l}\Big)\Big(1-\sum_{l=1}^{d-1} R_{\alpha,l}\Big)= \sum_{l=1}^{d} R_{\alpha,l}(1-R_{\alpha,l})=s_\alpha(say);\\
				\lVert \bm{M}\rVert_{sp}& = \lVert \Sigma_{\alpha,-d}+\Sigma_{\alpha,-d}^{1/2}\bm{1}\bm{1}^T\Sigma_{\alpha,-d}^{1/2} \rVert_{sp}\le \lVert \Sigma_{\alpha,-d}\rVert_{sp}+\bm{1}^T\Sigma_{\alpha,-d}\bm{1}\le \max\limits_{l=1,2,\ldots,d-1}R_{\alpha,l}+R_{\alpha,d}(1-R_{\alpha,d})=v_\alpha
			\end{aligned}
			$$
			as $\lVert \Sigma_{\alpha,-d}\rVert_{sp}\le\max\limits_{l=1,2,\ldots,d-1}R_{\alpha,l}$ by the result of \cite{watson1996spectral}. Hence,
			$$
			\begin{aligned}
				&P\Big(\lVert\hat{\boldsymbol\pi}_j-\bm{R}_\alpha\rVert_2\ge\dfrac{\epsilon}{2}\Big)\le \exp\Big[-\dfrac{a_2 ((n-m)\epsilon^2-4q_js_\alpha)}{4q_jv_\alpha}\Big]=\exp\Big(\dfrac{s_\alpha}{v_\alpha}\Big)\exp\Big[-\dfrac{a_2(n-m)\epsilon^2}{4q_jv_\alpha}\Big]\\
				\implies & P\Big(\lVert\hat{\boldsymbol\pi}_j-\bm{R}_\alpha\rVert_2<\dfrac{\epsilon}{2};\forall j\in\C_\alpha,\forall\alpha=1,\ldots,k_0\Big)\\
				\ge &1- \sum_{\alpha=1}^{k_0}\sum_{j\in\C_\alpha}P\Big(\lVert\hat{\boldsymbol\pi}_j-\bm{R}_\alpha\rVert_2\ge\dfrac{\epsilon}{2}\Big) \ge 1- \sum_{\alpha=1}^{k_0}\exp\Big(\dfrac{s_\alpha}{v_\alpha}\Big)\sum_{j\in\C_\alpha}\exp\Big[-\dfrac{a_2(n-m)\epsilon^2}{4q_jv_\alpha}\Big].
			\end{aligned}
			$$
			Now, setting  $C_1^{(\alpha)}=\exp\Big(\dfrac{s_\alpha}{v_\alpha}\Big)$, $C_{2,j}=\dfrac{a_2\epsilon^2}{4q_jv_\alpha}$, one 
			gets the conclusions of the theorem. 
			
			\subsection{Proof of Theorem \ref{thm_BIC}}\label{proof3}
			
			Recall that $\hpi_{j,\ell}= N_{\si_j,\ell}/N_{\si_j}$. Denote the common transition probability for the estimated group $\hat{\C}_\alpha(\lambda)$ as 
			$$
			\hat{R}^{(\lambda)}_{\alpha,\ell}=\dfrac{\sum_{\si_j\in\hat{\C}_\alpha(\lambda)}N_{\si_j,\ell}}{\sum_{\si_j\in\hat{\C}_\alpha(\lambda)}N_{\si_j}}=\dfrac{N_{\hat{\C}_\alpha(\lambda),\ell}}{N_{\hat{\C}_\alpha(\lambda)}}\quad\quad \forall \alpha=1,\ldots,k_{\lambda};\ell=1,\ldots,d.
			$$
			Thus, the log-likelihood is given by
			$$
			\ell_n(\lambda)=\sum_{\alpha=1}^{k_\lambda}\sum_{\ell=1}^{d} N_{\hat{\C}_\alpha(\lambda),\ell}\log \hat{R}^{(\lambda)}_{\alpha,\ell}.
			$$
			Note that, as $\lambda$ increases, the number of clusters decreases. Also, by the continuity of the solution of (\ref{cr1}) with respect to the $\lambda$, $M_{\lambda_2}$ is a submodel of $M_{\lambda_1}$ for $\lambda_1<\lambda_2,$ as the separate clusters for lower $\lambda$ values are clumped together to form new clusters as $\lambda$ increases. Hence, we can write $M_{\lambda_2}\subseteq M_{\lambda_1}$. Subsequently, $\ell_n(\lambda_1)\ge \ell_n(\lambda_2)$. Let $q_{j}$ be the stationary probability of the state $\si_j$, and let $Q^{(\alpha)}(\lambda)$ be the stationary probability of the partition $\hat{\C}_\alpha(\lambda)$. Thus, $Q^{(\alpha)}(\lambda)=\sum_{\si_j\in \hat{\C}_\alpha(\lambda)}q_j$. We have to show that the true model minimizes BIC with probability tending to $1$ as $n\to\infty$. We prove this for two cases.
			\\[.1in]
			\underline{Case 1:} Suppose that $\lambda<\lambda_0$ and $M_{\lambda_0}\subset M_{\lambda}$. Clearly, $k_{\lambda_0}<k_{\lambda}$. Since $M_{\lambda_0}$ is the true underlying model, $M_{\lambda_0}=\{\C_1,\ldots, \C_{k_0}\}$, and
			$$
			Z_n=-2\Big(\ell_n(\lambda_0)-\ell_n(\lambda)\Big) \xrightarrow{d} Z\sim\chi^2_{(k_{\lambda}-k_{0})(d-1)}.
			$$
			Hence, as $n\to\infty$,
			\beas
			P\Big(BIC_n(\lambda_0)\ge BIC_n(\lambda)\Big)&&=P\Big(Z_n>(k_{\lambda}-k_{0})(d-1)\log n\Big)\\
            && =P\Big(Z>(k_{\lambda}-k_{0})(d-1)\log n\Big) +\epsilon_n\\
			&&\le\exp\Big[-\dfrac{(k_{\lambda}-k_{0})(d-1)}{4} \log n \Big]+\epsilon_n\\ &&=n^{-\dfrac{(k_{\lambda}-k_{0})(d-1)}{4}} +\epsilon_n\to 0,
			\eeas
            where $\epsilon_n$ is the approximation error such that $\epsilon_n\to 0$ as $n\to\infty$. The inequality follows from a lemma by \cite{laurent2000adaptive}. If $Z\sim\chi^2_r$ distribution, then
            $$
            P(Z\ge r+2\sqrt{rx}+2x)\le e^{-x}.
            $$
            In our case, $r=(k_{\lambda}-k_{0})(d-1)$ and we want to solve for $x$ such that $r+2\sqrt{rx}+2x=r\log n$. By solving the quadratic equation, we get $x=\dfrac{r}{2}\Big(\log n-\sqrt{2\log n-1}\Big)\ge\dfrac{r \log n}{4}$ for large $n$. Hence the inequality holds.\\
			\underline{Case 2:} Now let $\lambda_0<\lambda$ and $M_{\lambda}\subset M_{\lambda_0}$. For $\alpha'=1,\ldots,k_\lambda$, without loss of generality, we can write
			$$
			\hat{\C}_{\alpha'}(\lambda)= \bigcup_{\alpha=t_{\alpha'-1}+1}^{\alpha=t_{\alpha'}} \C_\alpha
			$$
			for $0=t_0<t_1<t_2<\ldots<t_{k_\lambda}=k_0$.   Now, as $n\to\infty$,
			$$
			\begin{aligned}
				\dfrac{1}{n-m} \ell_n(\lambda_0)=\dfrac{1}{n-m} \sum_{\alpha=1}^{k_0}\sum_{\ell=1}^{d} N_{\C_\alpha,\ell}\log \hat{R}^{(\lambda_0)}_{\alpha,\ell}
				& \xrightarrow[]{p} \sum_{\alpha=1}^{k_0}\sum_{\ell=1}^{d} \Big(\sum_{j\in\C_\alpha} q_{\si_j}\Big)R_{\alpha,\ell}\log R_{\alpha,\ell}\\
				&=\sum_{\alpha=1}^{k_0}\sum_{\ell=1}^{d} Q^{(\alpha)}(\lambda_0)R_{\alpha,\ell}\log R_{\alpha,\ell}=A_0;
			\end{aligned}
			$$
			and
			
			$$
			\begin{aligned}
				\dfrac{1}{n-m} \ell_n(\lambda)&=\dfrac{1}{n-m} \sum_{\alpha'=1}^{k_\lambda}\sum_{\ell=1}^{d} N_{\hat{\C}_{\alpha'}(\lambda),\ell}\log \hat{R}^{(\lambda)}_{\alpha',\ell}
				=\dfrac{1}{n-m} \sum_{\alpha'=1}^{k_\lambda}\sum_{\ell=1}^{d} \Big(\sum_{j\in\hat{\C}_{\alpha'}(\lambda)} N_{\si_j,\ell}\Big) \log\Big(\dfrac{\sum_{j\in\hat{\C}_{\alpha'}(\lambda)} N_{\si_j,\ell}}{\sum_{j\in\hat{\C}_{\alpha'}(\lambda)} N_{\si_j}} \Big) \\
				&=\dfrac{1}{n-m}\sum_{\alpha'=1}^{k_\lambda}\sum_{\ell=1}^{d}\Big(\sum_{\alpha=t_{{\alpha'}-1}+1}^{t_{\alpha'}} N_{\C_\alpha,\ell}\Big)\log \Big(\dfrac{\sum_{\alpha=t_{{\alpha'}-1}+1}^{t_{\alpha'}} N_{\C_\alpha,\ell}}{\sum_{\alpha=t_{\alpha'-1}+1}^{t_{\alpha'}} N_{\C_\alpha}} \Big)\\
				& \xrightarrow[]{p} \sum_{\alpha'=1}^{k_\lambda}\sum_{\ell=1}^{d} \Big(\sum_{\alpha=t_{\alpha'-1}+1}^{t_{\alpha'}} Q^{(\alpha)}(\lambda_0)R_{\alpha,\ell}\Big)\log\Big(\dfrac{\sum_{\alpha=t_{\alpha'-1}+1}^{t_{\alpha'}} Q^{(\alpha)}(\lambda_0)R_{\alpha,\ell}}{\sum_{\alpha=t_{\alpha'-1}+1}^{t_{\alpha'}} Q^{(\alpha)}(\lambda_0)} \Big)=A(\lambda).
			\end{aligned}
			$$
			
			Now, applying Jensen's inequality by using the strict convexity of $-\log x$, 
			$$
			\begin{aligned}
				A(\lambda)&= -\sum_{\ell=1}^{d}\sum_{\alpha'=1}^{k_\lambda} \Big(\sum_{\alpha=t_{\alpha'-1}+1}^{t_{\alpha'}} Q^{(\alpha)}(\lambda_0)R_{\alpha,\ell}\Big)\log\Big(\dfrac{\sum_{\alpha=t_{\alpha'-1}+1}^{t_{\alpha'}} Q^{(\alpha)}(\lambda_0)} {\sum_{\alpha=t_{\alpha'-1}+1}^{t_{\alpha'}} Q^{(\alpha)}(\lambda_0)R_{\alpha,\ell}}\Big)\\
				&= -\sum_{\ell=1}^{d}\sum_{\alpha'=1}^{k_\lambda} \Big(\sum_{\alpha=t_{\alpha'-1}+1}^{t_{\alpha'}} Q^{(\alpha)}(\lambda_0)R_{\alpha,\ell}\Big)\log\Big(\dfrac{\sum_{\alpha=t_{\alpha'-1}+1}^{t_{\alpha'}} Q^{(\alpha)}(\lambda_0)R_{\alpha,\ell}.(1/R_{\alpha,\ell})} {\sum_{\alpha=t_{\alpha'-1}+1}^{t_{\alpha'}} Q^{(\alpha)}(\lambda_0)R_{\alpha,\ell}}\Big)\\
				& < -\sum_{\ell=1}^{d}\sum_{\alpha'=1}^{k_\lambda} \sum_{\alpha=t_{\alpha'-1}+1}^{t_{\alpha'}} Q^{(\alpha)}(\lambda_0)R_{\alpha,\ell} \log (1/R_{\alpha,\ell})\\
				&=\sum_{\ell=1}^{d}\sum_{\alpha'=1}^{k_\lambda} \sum_{\alpha=t_{\alpha'-1}+1}^{t_{\alpha'}} Q^{(\alpha)}(\lambda_0)R_{\alpha,\ell} \log R_{\alpha,\ell}=A_0.
			\end{aligned}
			$$
			Hence, $\dfrac{1}{n-m}(\ell_n(\lambda_0)- \ell_n(\lambda))\xrightarrow[]{p} A_0-A(\lambda)>0$, and $P\Big(\dfrac{1}{n-m}(\ell_n(\lambda_0)- \ell_n(\lambda))\ge\dfrac{1}{2}(A_0-A(\lambda)\Big)\to 1$ as $n\to\infty$. Since $\log n/N\to 0$ as $n\to \infty$,
			$$
			\begin{aligned}
				P\Big(BIC_n(\lambda_0)\ge BIC_n(\lambda) \Big)&=P\Big(2\ell_n(\lambda_0)\le 2\ell_n(\lambda) +(k_0-k_{\lambda})(d-1)\log n\Big)\\
				&=P\Big(\ell_n(\lambda_0)- \ell_n(\lambda)\le (k_0-k_{\lambda})(d-1)\log n\Big)\\
				&=P\Big(\frac{1}{n-m}(\ell_n(\lambda_0)- \ell_n(\lambda))\le (k_0-k_{\lambda})(d-1)\frac{\log n}{n-m}\Big)\\
				& \to 0. 
			\end{aligned}
			$$
			
			\subsection{Proof of Theorem \ref{thm_special_case}}\label{proof4}
			By definition, we can easily conclude that the weights $w_{i,j}$ are symmetric, hence the first part of (A1) is satisfied. Now, observe that,
			$$
			\begin{aligned}
				\lVert\hat{\boldsymbol\pi}_i-\hat{\boldsymbol\pi}_j\rVert_2 &\le \lVert\hat{\boldsymbol\pi}_i-\bm{R}_\alpha\rVert_2+ \lVert\hat{\boldsymbol\pi}_j-\bm{R}_\alpha\rVert_2 < \epsilon, \text{ for } i,j\in\mathcal{C}_{\alpha}\\
				\lVert\hat{\boldsymbol\pi}_i-\hat{\boldsymbol\pi}_j\rVert_2 & \ge \lVert \bm{R}_\alpha -\bm{R}_\beta \rVert_2 - \lVert\hat{\boldsymbol\pi}_i-\bm{R}_\alpha\rVert_2- \lVert\hat{\boldsymbol\pi}_j-\bm{R}_\beta\rVert_2\\
				& \ge \delta-\epsilon, \text{ for } i\in\mathcal{C}_{\alpha}, j\in\mathcal{C}_{\beta}, \alpha\neq\beta.
			\end{aligned}
			$$
			Hence, for $\epsilon<\delta/2$, $w_{i,j}>0$ for $i,j\in\mathcal{C}_{\alpha}$, and thus (A1) holds. \newline
			First, assume that the cluster sizes are different. Note that, for $i\in\mathcal{C}_{\alpha}$,
			$$
			\sum_{\beta\neq\alpha} w_i^{(\beta)}=\sum_{\beta\neq\alpha}\sum_{i'\in\mathcal{C}_\beta} w_{i,i'} \le (k'+1-p_{\alpha})\exp\big[-\phi(\delta-\epsilon)^2\big],
			$$
			since at most $k'-(p_{\alpha}-1)$ many $w_{i,i'}$ can be non-zero if $i'\notin \mathcal{C}_{\alpha}$. Thus, for $i,j\in\mathcal{C}_{\alpha}$
			$$
			\mu_{i,j}^{(\alpha)} \le \sum_{\beta\ne\alpha} w_i^{(\beta)} + \sum_{\beta\ne\alpha} w_j^{(\beta)} \le 2(k'+1-p_{\alpha})\exp\big[-\phi(\delta-\epsilon)^2\big];
			$$
			and
			$$
			\begin{aligned}
				\dfrac{\mu_{i,j}^{(\alpha)}}{p_{\alpha}w_{i,j}} & < \dfrac{2(k'+1-p_{\alpha})\exp\big[-\phi(\delta-\epsilon)^2\big] }{p_\alpha\exp\big[-\phi\epsilon^2\big]}  < 2\big(\dfrac{k'+1}{p_{min}}-1\big)\exp\big[-\phi(\delta^2-2\delta\epsilon)\big]\\
				&=\exp\big[2\phi\delta \big(\epsilon-\dfrac{\delta}{2}+\dfrac{1}{2\phi\delta}\log\Big(\frac{2(k'+1-p_{min})}{p_{min}}\Big)\big)\big]=\exp\big[2\phi\delta(\epsilon-\epsilon_{max})\big]\\
				&<1, \text{ for } \epsilon<\epsilon_{max}.
			\end{aligned}
			$$
			Thus Condition (A2) holds. Now,
			$$
			\begin{aligned}
				\delta_1 &\ge p_{min}\exp\big[-\phi\epsilon^2\big]-2(k'+1-p_{\alpha})\exp\big[-\phi(\delta-\epsilon)^2\big] \\
				&\ge p_{min}exp\big[-\phi\epsilon_{max}^2\big]-2(k'+1-p_{\alpha})\exp\big[-\phi(\delta-\epsilon_{max})^2\big]=\delta_1^{(min)}.
			\end{aligned}
			$$
			Also,
			$$
			w^{(\alpha,l)}=\sum_{i\in\mathcal{C}_{\alpha}}w_i^{(l)} \le p_\alpha (k'+1-p_{\alpha})\exp\big[-\phi(\delta-\epsilon)^2\big].
			$$
			Hence,
			$$
			\begin{aligned}
				\delta_2 &\le \max_{1\le\alpha<\beta\le k_0} (2k'+2-p_{\alpha}-p_{\beta})\exp\big[-\phi(\delta-\epsilon)^2\big] \\
				&\le 2(k'+1-p_{min})\exp\big[-\phi(\delta-\epsilon_{max})^2\big]=\delta_2^{(max)}. 
			\end{aligned}
			$$
			\subsubsection{Proof of Corollary \ref{cor_1}}
			Note that (a) follows directly from the definition of $\lambda_{\text{min}}^{(n)}$ and $\lambda_{\text{max}}^{(n)}$. Now if (a) holds, $\lambda_{\text{min}}^{(n)}<\lambda_{\text{max}}^{(n)}$ if $\epsilon<\dfrac{\delta\delta_1^{(min)}}{\delta_1^{(min)}+\delta_2^{(max)}}$, and (b) holds if $\epsilon<\epsilon_{max}$, proving the result.
			\subsubsection{Proof of Corollary \ref{cor_2}}
			Under the conditions of Theorem (\ref{thm_special_case}), $w_{i,j}>0$ for $i,j\in\mathcal{C}_{\alpha}$ for some $\alpha=1,\ldots,k_0$. If $p_{\alpha}=p/k_0$, and $k=p/k_0-1$, then $k'=k=p/k_0-1$ as well, since all the weights $w_{i,j}=0$ if the two $m$-tuples $\sigma_i$ and $\sigma_j$ belong to different cluster. Hence, $\delta_2^{(max)}=0$. Also, in this case $\mu_{i,j}^{(\alpha)}=0$ if $\epsilon<\delta/2$, and hence $p_{\alpha}w_{i,j}>\mu_{i,j}^{(\alpha)}=0$ for $i,j\in\mathcal{C}_{\alpha}$. Thus,
			$$
			\delta_1 \ge (p/k_0)\exp\big[-\phi\epsilon^2\big]\ge (p/k_0)\exp\big[-\phi(\delta/2)^2\big]
			$$
			and $\lambda_{\text{min}}^{(n)}<\dfrac{\delta}{2\delta_1}$, $\lambda_{\text{max}}^{(n)}=\infty$.

\section{Definitions}
\subsection{Adjusted Rand Index} \label{app:ari} Mathematically, for any two cluster assignments $X=(X_1,\ldots,X_r)$ and $Y=(Y_1,\ldots,Y_s)$ of the elements $(\si_1,\ldots,\si_p)$, the Rand Index is defined by 
$$
RI=\dfrac{a+b}{a+b+c+d}=\dfrac{a+b}{\binom{p}{2}} 
$$
where $a$ is the number of pairs that are in the same cluster in both $X$ and $Y$, $b$ is the number of pairs that are in the different clusters in both $X$ and $Y$, $c$ is the number of pairs that are in same cluster of $X$, but in different clusters of $Y$, and $d$ is number of pairs which are in same cluster of $Y$, but in different clusters of $X$. Values of $RI$ vary  between $0$ and $1$. If two clusters are identical, $RI$ should be $1$. Higher $RI$ values indicate more similarity among two given clusters.

However, the Rand Index has some limitations. For example, if the number of clusters increases, and the cluster sizes are not large, $RI$ will be close to $1$ even for two completely different cluster assignments. To address this issue, usage of the Adjusted Rand Index (ARI) is preferred. $ARI$ uses the expected similarity of all pairwise comparisons between clusterings specified by a random model. If $a_i=|X_i|$, $b_j=|Y_j|$, and $p_{ij}=|X_i\cap Y_j|$, then the $ARI$ is computed by the following formula:
$$
ARI=\dfrac{\sum\limits_{i,j}\binom{p_{ij}}{2}-\big[\sum\limits_{i}\binom{a_i}{2}\sum\limits_{j}\binom{b_j}{2}\big]/\binom{p}{2}}{\frac{1}{2}\big[\sum\limits_{i}\binom{a_i}{2}+\sum\limits_{j}\binom{b_j}{2}\big]-\big[\sum\limits_{i}\binom{a_i}{2}\sum\limits_{j}\binom{b_j}{2}\big]/\binom{p}{2}}.
$$
\section{Supplementary Tables}
\begin{table}[ht]
\centering
\resizebox{\textwidth}{!}{
\begin{tabular}{|c|c|c|cc|cc|cl|}
\hline
\multirow{2}{*}{\textbf{n}} & \multirow{2}{*}{\textbf{Method}}                                 & \multirow{2}{*}{\textbf{Uniform Weight}} & \multicolumn{2}{c|}{\textbf{Weight=}$l_2$}                        & \multicolumn{2}{c|}{\textbf{Weight=}$l_\infty$}                        & \multicolumn{2}{c|}{\multirow{2}{*}{\textbf{Weylandt}}} \\ \cline{4-7}
                            &                                                                  &                                          & \multicolumn{1}{c|}{\textbf{knn=5}} & \textbf{knn=3}         & \multicolumn{1}{c|}{\textbf{knn=5}} & \textbf{knn=3}         & \multicolumn{2}{c|}{}                                   \\ \hline
\multirow{2}{*}{5000}       & ARI (s.e.)                                                        & 0.044 (0.137)                            & \multicolumn{1}{c|}{0.745 (0.195)}  & 0.851 (0.171)          & \multicolumn{1}{c|}{0.720 (0.201)}  & \textbf{0.861 (0.170)} & \multicolumn{2}{c|}{0.776 (0.174)}                      \\ \cline{2-9} 
                            & \begin{tabular}[c]{@{}c@{}}Prob. of\\ True Recovery\end{tabular} & 0                                        & \multicolumn{1}{c|}{0.223}          & 0.48                   & \multicolumn{1}{c|}{0.187}          & \textbf{0.511}         & \multicolumn{2}{c|}{0.267}                              \\ \hline
\multirow{2}{*}{10000}      & ARI (s.e.)                                                        & 0.242 (0.187)                            & \multicolumn{1}{c|}{0.946 (0.102)}  & 0.983 (0.056)          & \multicolumn{1}{c|}{0.928 (0.117)}  & \textbf{0.984 (0.054)} & \multicolumn{2}{c|}{0.960 (0.083)}                      \\ \cline{2-9} 
                            & \begin{tabular}[c]{@{}c@{}}Prob. of\\ True Recovery\end{tabular} & 0                                        & \multicolumn{1}{c|}{0.708}          & 0.908                  & \multicolumn{1}{c|}{0.644}          & \textbf{0.908}         & \multicolumn{2}{c|}{0.784}                              \\ \hline
\multirow{2}{*}{15000}      & ARI (s.e.)                                                        & 0.413 (0.189)                            & \multicolumn{1}{c|}{0.982 (0.052)}  & 0.995 (0.028)          & \multicolumn{1}{c|}{0.974 (0.063)}  & \textbf{0.995 (0.028)} & \multicolumn{2}{c|}{0.985 (0.048)}                      \\ \cline{2-9} 
                            & \begin{tabular}[c]{@{}c@{}}Prob. of\\ True Recovery\end{tabular} & 0.001                                    & \multicolumn{1}{c|}{0.876}          & 0.972                  & \multicolumn{1}{c|}{0.821}          & \textbf{0.973}         & \multicolumn{2}{c|}{0.908}                              \\ \hline
\multirow{2}{*}{20000}      & ARI (s.e.)                                                        & 0.542 (0.191)                            & \multicolumn{1}{c|}{0.992 (0.037)}  & \textbf{0.998 (0.017)} & \multicolumn{1}{c|}{0.989 (0.044)}  & \textbf{0.998 (0.017)} & \multicolumn{2}{c|}{0.994 (0.033)}                      \\ \cline{2-9} 
                            & \begin{tabular}[c]{@{}c@{}}Prob. of\\ True Recovery\end{tabular} & 0.008                                    & \multicolumn{1}{c|}{0.951}          & \textbf{0.991}         & \multicolumn{1}{c|}{0.922}          & 0.99                   & \multicolumn{2}{c|}{0.964}                              \\ \hline
\multirow{2}{*}{25000}      & ARI (s.e.)                                                        & 0.656 (0.179)                            & \multicolumn{1}{c|}{0.997 (0.023)}  & \textbf{0.999 (0.011)} & \multicolumn{1}{c|}{0.994 (0.029)}  & \textbf{0.999 (0.011)} & \multicolumn{2}{c|}{0.997 (0.025)}                      \\ \cline{2-9} 
                            & \begin{tabular}[c]{@{}c@{}}Prob. of\\ True Recovery\end{tabular} & 0.028                                    & \multicolumn{1}{c|}{0.977}          & \textbf{0.996}         & \multicolumn{1}{c|}{0.96}           & \textbf{0.996}         & \multicolumn{2}{c|}{0.981}                              \\ \hline
\end{tabular}
}
\caption{Mean ARI (Standard Deviation) and True Recovery for different weight choices in convex clustering in Simulation 1.}
\label{tab:sim_1_cvx}
\end{table}

\begin{table}[ht]
\centering
\resizebox{\textwidth}{!}{
\begin{tabular}{|c|c|cc|c|c|c|cl|}
\hline
\multirow{2}{*}{\textbf{n}} & \multirow{2}{*}{\textbf{Method}}                                 & \multicolumn{2}{c|}{\textbf{Convex Clustering}}                      & \multicolumn{1}{l|}{\multirow{2}{*}{\textbf{GSDPMM}}} & \multicolumn{1}{l|}{\multirow{2}{*}{\textbf{Xiong}}} & \multicolumn{1}{l|}{\multirow{2}{*}{\textbf{GGL}}} & \multicolumn{2}{l|}{\multirow{2}{*}{\textbf{K-means}}} \\ \cline{3-4}
                            &                                                                  & \multicolumn{1}{c|}{knn=3, weight=$l_2$}    & knn=3, weight=$l_2$    & \multicolumn{1}{l|}{}                                 & \multicolumn{1}{l|}{}                                & \multicolumn{1}{l|}{}                              & \multicolumn{2}{l|}{}                                  \\ \hline
\multirow{2}{*}{5000}       & ARI (s.e.)                                                        & \multicolumn{1}{c|}{0.851 (0.171)}          & \textbf{0.861 (0.17)}  & 0.845 (0.167)                                         & 0.82 (0.172)                                         & 0.413 (0.136)                                      & \multicolumn{2}{c|}{0.703 (0.216)}                     \\ \cline{2-9} 
                            & \begin{tabular}[c]{@{}c@{}}Prob. of\\ True Recovery\end{tabular} & \multicolumn{1}{c|}{0.48}                   & \textbf{0.511}         & 0.451                                                 & 0.393                                                & 0.003                                              & \multicolumn{2}{c|}{0.187}                             \\ \hline
\multirow{2}{*}{10000}      & ARI (s.e.)                                                        & \multicolumn{1}{c|}{0.983 (0.056)}          & \textbf{0.984 (0.054)} & 0.976 (0.066)                                         & 0.952 (0.105)                                        & 0.706 (0.18)                                       & \multicolumn{2}{c|}{0.824 (0.201)}                     \\ \cline{2-9} 
                            & \begin{tabular}[c]{@{}c@{}}Prob. of\\ True Recovery\end{tabular} & \multicolumn{1}{c|}{\textbf{0.908}}         & \textbf{0.908}         & 0.876                                                 & 0.799                                                & 0.153                                              & \multicolumn{2}{c|}{0.357}                             \\ \hline
\multirow{2}{*}{15000}      & ARI (s.e.)                                                        & \multicolumn{1}{c|}{0.995 (0.028)}          & \textbf{0.995 (0.028)} & 0.994 (0.033)                                         & 0.981 (0.069)                                        & 0.876 (0.139)                                      & \multicolumn{2}{c|}{0.825 (0.199)}                     \\ \cline{2-9} 
                            & \begin{tabular}[c]{@{}c@{}}Prob. of\\ True Recovery\end{tabular} & \multicolumn{1}{c|}{0.972}                  & \textbf{0.973}         & 0.963                                                 & 0.914                                                & 0.494                                              & \multicolumn{2}{c|}{0.355}                             \\ \hline
\multirow{2}{*}{20000}      & ARI (s.e.)                                                        & \multicolumn{1}{c|}{\textbf{0.998 (0.017)}} & \textbf{0.998 (0.017)} & 0.997 (0.024)                                         & 0.992 (0.043)                                        & 0.943 (0.099)                                      & \multicolumn{2}{c|}{0.834 (0.194)}                     \\ \cline{2-9} 
                            & \begin{tabular}[c]{@{}c@{}}Prob. of\\ True Recovery\end{tabular} & \multicolumn{1}{c|}{\textbf{0.991}}         & 0.99                   & 0.982                                                 & 0.956                                                & 0.735                                              & \multicolumn{2}{c|}{0.367}                             \\ \hline
\multirow{2}{*}{25000}      & ARI (s.e.)                                                        & \multicolumn{1}{c|}{\textbf{0.999 (0.011)}} & \textbf{0.999 (0.011)} & 0.999 (0.014)                                         & 0.996 (0.029)                                        & 0.975 (0.067)                                      & \multicolumn{2}{c|}{0.819 (0.199)}                     \\ \cline{2-9} 
                            & \begin{tabular}[c]{@{}c@{}}Prob. of\\ True Recovery\end{tabular} & \multicolumn{1}{c|}{\textbf{0.996}}         & \textbf{0.996}         & 0.994                                                 & 0.978                                                & 0.872                                              & \multicolumn{2}{c|}{0.323}                             \\ \hline
\end{tabular}
}
\caption{Mean ARI (Standard Deviation) and True Recovery for different methods of fitting SMM for Simulation 1.}
\label{tab:sim_1_comparison}
\end{table}

		
		
		
		
		

		\bibliographystyle{plainnat}
		
		\bibliography{reference}
	\end{document}